\documentclass[
 reprint,
 superscriptaddress,
 amsmath,amssymb,
 aps,prx,
]{revtex4-1}

\usepackage{physics}

\usepackage{braket}
\usepackage{bm}
\usepackage{graphicx}
\usepackage{enumitem}
\usepackage[colorlinks]{hyperref}
\hypersetup{
citecolor=blue
}

\newcommand{\EDIT}[1]{{\color{black}#1}}

\graphicspath{ {./figures/} }

\begin{document}
\title{Maximally-Localized Exciton Wannier Functions for Solids}

\author{Jonah B. Haber}
\affiliation{Department of Physics, University of California Berkeley, Berkeley, California 94720, USA}

\author{Diana Y. Qiu}
\affiliation{Department of 	
Mechanical Engineering and Materials Science, Yale University, New Haven, Connecticut 06520}

\author{Felipe H. da Jornada}
\affiliation{Department of Materials Science and Engineering, Stanford University, Stanford, CA 94305, USA}

\author{Jeffrey B. Neaton}
\affiliation{Department of Physics, University of California Berkeley, Berkeley, California 94720, USA}
\affiliation{Materials Sciences Division, Lawrence Berkeley National Laboratory, Berkeley, California 94720, USA}
\affiliation{Kavli Energy Nanosciences Institute, Berkeley, California 94720, USA}

\begin{abstract}
We introduce a maximally-localized Wannier function representation of Bloch excitons, two-particle correlated electron-hole excitations, in crystalline solids, where the excitons are maximally-localized with respect to an average electron-hole coordinate in real space. As a proof-of-concept, we illustrate this representation in the case of low-energy spin-singlet and triplet excitons in LiF, computed using the \textit{ab initio} Bethe-Salpeter equation approach. We visualize the resulting maximally-localized exciton Wannier functions (MLXWFs) in real space, detail the convergence of the exciton Wannier spreads, and demonstrate how Wannier-Fourier interpolation can be leveraged to obtain exciton energies and states at arbitrary exciton crystal momenta in the Brillouin zone. We further introduce an approach to treat the long-range dipolar coupling between singlet MLXWFs and discuss it in depth. The MLXWF representation sheds light on the fundamental nature of excitons and paves the way towards Wannier-based post-processing of excitonic properties, enabling the construction of \textit{ab initio} exciton tight-binding models, efficient interpolation of the exciton-phonon vertex, the computation of Berry curvature associated with exciton bands, and beyond.
\end{abstract}

\maketitle

\section{\label{sec:intro} Introduction}

Since their introduction 25 years ago~\cite{Marzari1997-kz}, maximally-localized Wannier functions (MLWFs) have had a transformative impact on our ability to compute and understand one-electron observables using density functional theory (DFT). Today MLWFs serve as a compact basis for linear scaling algorithms~\cite{Williamson2001-ma}, allow for the computation of Berry phase (geometric) quantities~\cite{Wang2006-hc,Ibanez-Azpiroz2018-vv,Tsirkin2018-xk} (e.g. electronic polarization~\cite{King-Smith1993-vi}), and find application in efficient and accurate interpolation of linear response quantities (e.g. electron-phonon matrix elements)~\cite{Giustino2007-kc,Noffsinger2010-nq,Ponce2016-sg}, and more~\cite{Marzari2012-qr}.

In the MLWF scheme, periodic Bloch states are related to localized Wannier functions through a unitary transformation, one which simultaneously preserves the canonical commutation relations while also localizing the sum of the spreads of the states to the greatest extent possible~\cite{Marzari1997-kz,Souza2001-xr,Marzari2012-qr}. While the scheme is most often applied to one-electron Bloch states, the MLWF procedure is general and can be applied to any lattice periodic function. Indeed this framework has been used to construct localized representations of lattice vibrations~\cite{Rabe1995-av}, the electromagnetic field in photonic crystals~\cite{Garcia-Martin2003-kh} and, recently, perturbations to the electronic wavefunction~\cite{Lihm2021-mf}. 

In principle, the scheme can formally be applied to multi-particle states that are Bloch periodic in some average coordinate~\cite{Kohn1964-qu,Koch2001-my,Souza2000-hp}. One such excitation of broad theoretical and technological interest is the exciton, a two-particle \EDIT{correlated} electron-hole state. To date, localized representations of the exciton have only been constructed in the isolated limit~\cite{Tamura2016-pj} and in one dimension~\cite{Jin2017-kb}. Here, we introduce a general and practical extension of the MLWF scheme for excitons in a crystalline solid which may also serve as a blueprint for other multi-particle states.

Excitons are \EDIT{correlated} electron-hole pairs which often dominate the low-energy optical response of semiconducting and insulating materials. Understanding these composite particles plays an increasingly important role in the design and development of next-generation of optoelectronic devices, especially those based on complex materials with strong light-matter interactions. Over the past two decades, \textit{ab initio} many-body perturbation theory within the $GW$ approximation, where $G$ is the one-electron Green's function and $W$ the screened Coulomb interaction, and \EDIT{the} Bethe-Salpeter equation (BSE) approach~\cite{Hedin1965-hy, Hedin1970-xr, Strinati1988-wy, Hybertsen1986-bh, Rohlfing2000-wx} has rapidly emerged as a powerful and robust method for predicting excitonic properties for a wide range of increasingly complex materials including low-dimensional transition metal dichalcogenides \cite{Qiu2013-oq}, lead-halide perovskites \cite{Zhu2014-zr,Molina-Sanchez2018-ws,Biega2021-pj}, and organic crystals \cite{Tiago2003-ui,Sharifzadeh2012-xw,Sharifzadeh2013-ys,Rangel2016-ve}, in all cases yielding results in \EDIT{good} agreement with experiment. Given the technological relevance and increasing maturity of computational architectures and algorithms~\cite{Deslippe2012-bq}, revisiting the MLWF scheme in the context of excitons is timely.

In solid-state systems with translational symmetry, the exciton wavefunction can be written in Bloch periodic form with respect to an average electron-hole coordinate. A canonical example is the phenomenological Mott-Wannier model~\cite{Wannier1937-ea,Mott1938-aw} of a weakly-bound exciton; here, the exciton is described by a hydrogen-like wavefunction with the form
\begin{equation} \label{eq:mott_wann_xct}
    \Psi_{n\ell m,\mathbf{Q}}(\mathbf{R,\mathbf{r}}) = e^{i\mathbf{Q}\cdot \mathbf{R}}F_{n\ell m}(\mathbf{r}),
\end{equation}
where $\mathbf{R}$ and $\mathbf{r}$ denote the average and relative coordinate of the electron-hole pair, respectively, while $F_{n\ell m}(\mathbf{r})$ is a hydrogenic wavefunction with quantum numbers $n\ell m$. Notably, $\Psi_{n\ell m,\mathbf{Q}}(\mathbf{R,\mathbf{r}})$ obeys Bloch's theorem in $\mathbf{R}$ -- i.e. $\Psi_{\mathbf{Q},n\ell m}(\mathbf{R}+\bar{\mathbf{R}},\mathbf{r}) = e^{i\mathbf{Q}\cdot \bar{\mathbf{R}}}\Psi_{\mathbf{Q},n\ell m}(\mathbf{R},\mathbf{r})$, where $\bar{\mathbf{R}}$ is a lattice vector -- a general feature rigorously true of all descriptions of an exciton in a perfect crystal, e.g. \EDIT{excitons classified as} Frenkel~\cite{Frenkel1931-qi} \EDIT{or} charge-transfer \EDIT{in crystals} also obey Bloch's theorem in $\mathbf{R}$.

In this work we advance a new representation of the exciton in a periodic crystal, one which is maximally localized in an average electron-hole coordinate. This new representation is a natural but as-yet unexplored extension of one-electron Wannier functions to two-particle excitations and provides a rigorous particle-like picture of the exciton complimentary to the usual wave-like picture afforded by the Bloch representation of an exciton in a periodic solid.

The maximally-localized exciton Wannier functions (MLXWFs) allow for post-processing of exciton related properties in analogy to the electronic case, for instance, the \textit{ab initio} construction of exciton tight-binding models, the efficient interpolation of exciton eigenenergies~\cite{Qiu2015-bm,Qiu2021-ll} and exciton-phonon matrix elements~\cite{Toyozawa1958-dq,Antonius2022-lh,Chen2020-cm} throughout the Brillouin zone, the computation of Berry curvature related properties at the excitonic level~\cite{Kozin2021-sd,Onga2017-tk,Kuga2008-jw,Yao2008-gj}, and more. Our framework also \EDIT{deepens our understanding of} the phenomenological Mott-Wannier~\cite{Wannier1937-ea,Mott1938-aw} and Frenkel~\cite{Frenkel1931-qi} descriptions of an exciton in periodic solids. It further provides a physical picture for the splitting of the transverse and longitudinal exciton branches of the exciton band structure~\cite{Heller1951-vk, Hopfield1960-fa} in terms of long-range dipole-dipole interactions between MLXWFs.

The reminder of this paper is organized as follows. In Sec.~\ref{sec:xwf} we introduce exciton Wannier functions. In Sec.~\ref{sec:3} we review the MLWF framework and adapt this formalism for excitons. In Sec.~\ref{sec:4} we focus on matrix elements of the exciton Hamiltonian in the MLXWF basis. We show singlet excitons are coupled over large distances through a dipolar interaction, and we draw an analogy to the theory of lattice vibrations in a hetero-polar crystal. In Sec.~\ref{sec:5} we detail how to effectively partition the singlet exciton Hamiltonian into a short-range part, amenable to Wannier-based post-processing, and a long-range part which can be treated analytically. In Sec.~\ref{sec:6} we detail our Wannier-Fourier interpolation scheme for obtaining exciton band structures throughout the Brillouin zone. In Sec.~\ref{sec:7} we summarize the $GW$-BSE approach and computational details of our calculation. In Sec.~\ref{sec:8} we apply our framework to cubic LiF. We tabulate the convergence of the Wannier spreads, visualize the MLXWFs in real-space, and show the results of our Wannier-Fourier interpolation for the lowest three exciton bands. We draw comparisons for both the singlet and triplet cases. In Sec.~\ref{sec:9} we give a discussion detailing potential future applications of MLXWFs. We close in Sec.~\ref{sec:10} with a summary of our work.  

\section{Exciton Wannier Function \label{sec:xwf}} 

An exciton is a composite quasiparticle consisting of a \EDIT{correlated} electron-hole pair. The position of the electron, $\mathbf{r}_e$, and hole, $\mathbf{r}_h$, are correlated and encoded in the exciton wavefunction, $\Psi_{S\mathbf{Q}}(\mathbf{r}_e,\mathbf{r}_h)$, where $S$ and $\mathbf{Q}$ denote the exciton's principle quantum number and crystal momentum, respectively. Physically, $\Psi_{S\mathbf{Q}}(\mathbf{r}_e,\mathbf{r}_h)$ is the probability amplitude to simultaneously find an electron and hole at $\mathbf{r}_e$ and $\mathbf{r}_h$, respectively.

An increasingly standard approach for computing exciton states and properties in solids is the \textit{ab initio} $GW$-BSE method, mentioned above. In this approach, $\Psi_{S\mathbf{Q}}(\mathbf{r}_e,\mathbf{r}_h)$ is expressed as a coherent sum over non-interacting electron-hole product states, namely,
\begin{equation} \label{eq:bse_exciton}
\Psi_{S\mathbf{Q}}(\mathbf{r}_e,\mathbf{r}_h) =  \sum_{cv\mathbf{k}} A^{S\mathbf{Q}}_{cv\mathbf{k}} \psi_{c\mathbf{k}}(\mathbf{r}_e) \psi^\star_{v\mathbf{k}-\mathbf{Q}}(\mathbf{r}_h),
\end{equation}
where $\psi_{n\mathbf{k}}(\mathbf{r})= e^{i \mathbf{k} \cdot \mathbf{r}}u_{n\mathbf{k}}(\mathbf{r})$ denotes a single-particle Bloch state with band index $n$ and crystal momentum $\mathbf{k}$ (typically computed from Kohn-Sham DFT, Hartree Fock etc.), while $A_{cv\mathbf{k}}^{S\mathbf{Q}}$ is the exciton expansion coefficient with subscript $c$ ($v$) indexing conduction (valence) states. The same exciton wavefunction can be written in Bloch periodic form when re-expressed in the average $\mathbf{R}=(\mathbf{r}_e+\mathbf{r}_h)/2$ and relative $\mathbf{r} = \mathbf{r}_e - \mathbf{r}_h$ coordinates. Explicitly
\begin{equation} \label{eq:bloch_xct}
\begin{split}
\Psi_{S\mathbf{Q}}(\mathbf{R},\mathbf{r})  = \frac{1}{\sqrt{N_{\mathbf{Q}}}} e^{i\mathbf{Q}\cdot \mathbf{R}}F_{S\mathbf{Q}}(\mathbf{R},\mathbf{r}),
\end{split}
\end{equation}
where $F_{S\mathbf{Q}}(\mathbf{R},\mathbf{r})$ is given by
\begin{equation} \label{eq:FSQ}
\begin{split}
F_{S\mathbf{Q}}(\mathbf{R},\mathbf{r}) =  &\frac{1}{\sqrt{N_{\mathbf{k}}}} \sum_{cv\mathbf{k}} A^{S\mathbf{Q}}_{cv\mathbf{k}+\mathbf{Q}/2} e^{i\mathbf{k} \cdot \mathbf{r}} \\
&\times u_{c\mathbf{k}+\EDIT{\mathbf{Q}/2}}(\mathbf{R}+\mathbf{r}/2) u^\star_{v\mathbf{k}-\EDIT{\mathbf{Q}/2}}(\mathbf{R}-\mathbf{r}/2).
\end{split}
\end{equation}
Importantly $F_{S\mathbf{Q}}(\mathbf{R},\mathbf{r})$ is cell-periodic in $\mathbf{R}$ but not in $\mathbf{r}$. (Note Eq.~\ref{eq:mott_wann_xct} is a specialized case of this general form.) 

For present purposes, we assume that the exciton states are computed on regular $\mathbf{Q}$- and $\mathbf{k}$-meshes with $N_{\mathbf{Q}}$ and $N_{\mathbf{k}}$ points, respectively. Our exciton wavefunctions are normalized such that 
\begin{equation}
    \int |\Psi_{S\mathbf{Q}}(\mathbf{R},\mathbf{r})|^2 d\mathbf{r} d\mathbf{R} = 1,
\end{equation}
where the integrals in $\mathbf{R}$ and $\mathbf{r}$ are to be performed over supercells with volume $N_{\mathbf{Q}}V_{\text{uc}}$ and  $N_{\mathbf{k}}V_{\text{uc}}$, respectively, with $V_{\text{uc}}$ denoting the volume of the unit cell. From here on all integrals in $\mathbf{R}$ and $\mathbf{r}$ are understood to be taken over their respective supercell volumes unless otherwise stated. In Appendix~\ref{app:A}, we provide additional details on this change of coordinates and our normalization conventions. In the same Appendix we also discuss a generalization where the exciton is expressed in a weighted-average coordinate $\mathbf{R} = \alpha \mathbf{r}_e+\beta \mathbf{r}_h$, with $\alpha,\beta \ge 0$ and $\alpha+\beta=1$. For the sake of conceptual simplicity, we will continue to work with the specialized case $\alpha=\beta=1/2$ throughout the main text.

In this work, we define the exciton Wannier function as a Wannier transform of the Bloch exciton in the average electron-hole coordinate, explicitly
\begin{equation} \label{eq:xct_wann}
    W_{S\bar{\mathbf{R}}}(\mathbf{R},\mathbf{r}) = \frac{1}{\sqrt{N_{\mathbf{Q}}}} \sum_{\mathbf{Q}} e^{-i\mathbf{Q}\cdot \bar{\mathbf{R}}} \Psi_{S\mathbf{Q}}(\mathbf{R},\mathbf{r}), 
\end{equation}
where $\bar{\mathbf{R}}$ is a lattice vector. The exciton Wannier functions span the same functional space as the original excitonic states and by construction are orthonormal in indices $(S,\bar{\mathbf{R}})$ so that $\braket{W_{S\bar{\mathbf{R}}}|W_{S'\bar{\mathbf{R}'}}} =  \delta_{SS'}\delta_{\bar{\mathbf{R}}\bar{\mathbf{R}}'}$.  When translated by a lattice vector, they transform as $W_{S\bar{\mathbf{R}}}(\mathbf{R}-\bar{\mathbf{R}}',\mathbf{r}) = W_{S\bar{\mathbf{R}}+\bar{\mathbf{R}}'}(\mathbf{R},\mathbf{r})$, a result which immediately follows from Bloch's theorem. 

Further, the exciton Wannier functions are localized in the coordinate $\mathbf{R}$ and reside in the cell labeled by $\bar{\mathbf{R}}$. This is readily seen in the limit where the cell-periodic function is only weakly dependent on $\mathbf{Q}$ so that $F_{S\mathbf{Q}}(\mathbf{R},\mathbf{r}) \approx F_{S}(\mathbf{R},\mathbf{r})$. Then, $ W_{S\bar{\mathbf{R}}}(\mathbf{R},\mathbf{r}) \approx F_S(\mathbf{R},\mathbf{r}) \sum_{\mathbf{Q}} e^{i\mathbf{Q}\cdot (\mathbf{R}-\bar{\mathbf{R}})}$. Since the only length scales appearing in the sum are the lattice parameters, $W_{S\bar{\mathbf{R}}}(\mathbf{R},\mathbf{r})$ should decay rapidly in $\mathbf{R}-\bar{\mathbf{R}}$ beyond a few multiples of $\bar{\mathbf{R}}$. By contrast, the spread in the relative coordinate, $\mathbf{r}$, is related to the exciton radius, which, for weakly-bound excitons, can be on the order of many unit cells. We expect there are many scenarios where the spread in the average coordinate is smaller than the spread in the relative coordinate.

\section{Maximally-Localized Exciton Wannier Functions \label{sec:3}} 

In the MLWF scheme, extended one-electron Bloch states, $\psi_{n\mathbf{k}}(\mathbf{r}_e) = \braket{\mathbf{r}_e|n\mathbf{k}}$, are related to localized Wannier functions with orbital index $m$, $w_{m\bar{\mathbf{R}}}(\mathbf{r}_e)= \braket{\mathbf{r}_e|m\bar{\mathbf{R}}}$, through a unitary transformation
\begin{equation} \label{eq:el_wann}
w_{m\bar{\mathbf{R}}}(\mathbf{r}_e) = \frac{1}{\sqrt{N_{\mathbf{k}}}} \sum^{N_{\EDIT{\mathcal{W}}}}_{n\mathbf{k}} e^{-i\mathbf{k} \cdot \bar{\mathbf{R}}} U_{nm}(\mathbf{k})  \psi_{n\mathbf{k}}(\mathbf{r}_e),
\end{equation}
where $U_{nm}(\mathbf{k})$ is a unitary matrix which mixes some subset of states, $n\in \EDIT{\mathcal{W}}$, at a given $\mathbf{k}$-point. 

In developing this framework, Marzari and Vanderbilt~\cite{Marzari1997-kz} took advantage of the extra gauge freedom in $U_{nm}(\mathbf{k})$ to localize the sum of the spread of the Wannier functions to the greatest extent possible. For MLWFs, the sum of the spreads is defined as
\begin{equation} \label{eq:tot_spread}
\Omega^{\text{el}}[U] = \sum_m^{N_{\EDIT{\mathcal{W}}}} [ \braket{m\bar{\mathbf{0}}|r_e^2|m\bar{\mathbf{0}}} - \braket{m\bar{\mathbf{0}}|\mathbf{r}_e|m\bar{\mathbf{0}}}^2],    
\end{equation}
where $\mathbf{r}_e$ is understood as the position operator and the notation, $\Omega^{\text{el}}[U]$, indicates that the spread is a functional of the gauge, $U$.

Nowadays the minimization procedure is often performed using \texttt{Wannier90}, an open source, post-processing software for constructing MLWFs, compatible with many DFT codes~\cite{Mostofi2014-jy, Pizzi2020-ff}. 
\texttt{Wannier90} minimizes Eq.~\ref{eq:tot_spread} in reciprocal space and requires as a key input the overlaps of Bloch periodic states at neighboring \textbf{k}-points, $M_{nm}^{(\mathbf{k},\mathbf{b})}=\braket{u_{n\mathbf{k}}|u_{m\mathbf{k}}}_{\text{uc}}$, where the subscript ``uc", indicates that the overlaps are to be computed in the unit cell.  As emphasized by the developers of~\texttt{Wannier90}, these inputs are entirely agnostic to the underlying electronic structure theory calculation~\cite{Mostofi2014-jy, Pizzi2020-ff}. 

A more subtle point is that MLWF procedure itself is agnostic to the type of quasiparticle excitation which one wishes to localize so long as the excitation can be written in Bloch form. Said another way, the procedure can be used to find localized representations of any lattice periodic excitation. Accordingly, in analogy with Eq.~\ref{eq:el_wann}, exciton states, $\Psi_{S\mathbf{Q}}(\mathbf{R},\mathbf{r}) = \braket{\mathbf{R}\mathbf{r}|S\mathbf{Q}}$, are also related to localized exciton Wannier functions, $W_{M\bar{\mathbf{R}}}(\mathbf{R},\mathbf{r}) =\braket{\mathbf{R}\mathbf{r}|M\bar{\mathbf{R}}}$, through a unitary transformation, namely
\begin{equation} \label{eq:xct_mlwf}
   W_{M\bar{\mathbf{R}}}(\mathbf{R},\mathbf{r}) = \frac{1}{\sqrt{N_{\mathbf{Q}}}} \sum_{S\mathbf{Q}} e^{-i\mathbf{Q}\cdot \bar{\mathbf{R}}} U_{SM}(\mathbf{Q}) \Psi_{S\mathbf{Q}}(\mathbf{R},\mathbf{r}),
\end{equation}
where $M$ denotes the principal quantum number of the exciton Wannier function. 

In further analogy with the one-electron case, we can leverage the gauge freedom and choose $U_{MS}(\mathbf{Q})$ to minimize the sum of spread of the exciton Wannier functions defined as
\begin{equation} \label{eq:xct_tot_spread}
\Omega^{\text{Xct}}[U] = \sum_M^{N_{\EDIT{\mathcal{W}}}} [ \braket{M\bar{\mathbf{0}}|R^2|M\bar{\mathbf{0}}} - \braket{M\bar{\mathbf{0}}|\mathbf{R}|M\bar{\mathbf{0}}}^2],
\end{equation}
with
\begin{equation} \label{eq:xct_msd}
    \braket{M\bar{\mathbf{0}}|R^2|M\bar{\mathbf{0}}} = \int |W_{M\bar{\mathbf{0}}}(\mathbf{R},\mathbf{r})|^2 R^2 d \mathbf{r} d \mathbf{R}.
\end{equation}
As emphasized previously, the spread here is with respect to the average, \EDIT{and} not the relative\EDIT{,} coordinate. 

We use the \texttt{Wannier90} package to minimize Eq.~\ref{eq:xct_tot_spread}. In practice this is done by passing overlap matrices, $M_{SS'}(\mathbf{Q},\mathbf{B}) = \braket{F_{S\mathbf{Q}}|F_{S'\mathbf{Q}+\mathbf{B}}}$, to \texttt{Wannier90}. We compute these overlaps in the electron-hole basis as follows
\begin{equation} \label{eq:M_SS}
\begin{split}
    M_{SS'}&(\mathbf{Q},\mathbf{B}) = \sum_{cc'vv'\mathbf{k}}  [A^{S\mathbf{Q}}_{cv\mathbf{k}}]^\star A^{S'\mathbf{Q}+\mathbf{B}}_{c'v'\mathbf{k}+\mathbf{B}/2}  \\
    &\times \braket{u_{c\mathbf{k}}|u_{c'\mathbf{k}+\mathbf{B}/2}}_{\text{uc}} \braket{u_{v'\mathbf{k}-\mathbf{Q}-\mathbf{B}/2}|u_{v\mathbf{k}-\mathbf{Q}}}_{\text{uc}}.
\end{split}
\end{equation}
Eq.~\ref{eq:M_SS} constitutes a pivotal result of this work. Importantly it allows us to compute overlaps between the cell-periodic part of exciton wavefunctions with quantities readily obtained from standard electronic software codes. A generalization of Eq.~\ref{eq:M_SS} for arbitrary $\mathbf{R}=\alpha\mathbf{r}_e+\beta\mathbf{r}_h$ is derived in Appendix~\ref{app:B}, \EDIT{see Eq.~\ref{eq:MSS_derivation} specifically}.

In Table~\ref{tab:comparison}, we distinguish the notation used for, and draw an analogy between, electron and exciton Wannier functions.

\begin{table}[!ht] 
\centering
\begin{tabular}{c c c } 
 \hline \hline
  & electron & exciton \\ 
  \hline
 Bloch state & $\psi_{n\mathbf{k}}(\mathbf{r}_e)$ & $\Psi_{S\mathbf{Q}}(\mathbf{R},\mathbf{r})$   \\ 
 Wannier coordinate & $\mathbf{r}_e$ & $\mathbf{R}$ \\
 Conjugate momentum & $\mathbf{k}$ & $\mathbf{Q}$ \\
 Conjugate position & $\bar{\mathbf{R}}$ & $\bar{\mathbf{R}}$ \\
 Wannier function & $w_{m\bar{\mathbf{R}}}(\mathbf{r}_e)$ & $W_{M\bar{\mathbf{R}}}(\mathbf{R},\mathbf{r})$   \\ 
 Rotation matrix & $U_{nm}(\mathbf{k})$ & $U_{SM}(\mathbf{Q})$ \\
 \hline
 \hline
\end{tabular} 
\caption{Notation for, and analogy between, electron and exciton Wannier functions and related parameters. As defined in the text, $\mathbf{r}=\mathbf{r}_e-\mathbf{r}_h$ and $\mathbf{R}=\alpha\mathbf{r}_e+\beta\mathbf{r}_h$, with $\alpha=\beta=1/2$ being used throughout this work (see Appendix~\ref{app:A}).} \label{tab:comparison}
\end{table}

For use in the next section, it is convenient to note that Eq.~\ref{eq:xct_mlwf} can be inverted to give
\begin{equation} \label{eq:xct_mlwf_inv}
   \Psi_{S\mathbf{Q}}(\mathbf{R},\mathbf{r})  = \frac{1}{\sqrt{N_{\mathbf{Q}}}}\sum_{M\bar{\mathbf{R}}} e^{i\mathbf{Q}\cdot \bar{\mathbf{R}}} U^{\dagger}_{MS}(\mathbf{Q}) W_{M\bar{\mathbf{R}}}(\mathbf{R},\mathbf{r}).
\end{equation}
Eq.~\ref{eq:xct_mlwf_inv} can be derived from Eq.~\ref{eq:xct_mlwf}, with the use of the identity $\sum_\mathbf{\bar{\mathbf{R}}} e^{i(\mathbf{Q}-\mathbf{Q}')\cdot\bar{\mathbf{R}}} =N_{\mathbf{Q}}\delta_{\mathbf{Q}\mathbf{Q}'}$.

\section{Analytic properties of the exciton Hamiltonian \label{sec:4}}

An important property of one-electron MLWFs is that they form a compact basis for expanding the local one-electron Hamiltonian. In this case locality of the MLWFs and one-electron Hamiltonian guarantees that matrix elements in the MLWF basis -- i.e.,  $\braket{m\bar{\mathbf{0}}|H^{\text{el}}|m\bar{\mathbf{R}}}$ -- decay rapidly with increasing $\bar{\mathbf{R}}$. The same is not immediately apparent for the exciton Hamiltonian which contains non-local interactions that couple the electron and hole degrees of freedom. 

In the absence of spin-orbit coupling, we can partition the exciton Hamiltonian into a spin-singlet and spin-triplet sector. In this total spin basis, the exciton Hamiltonian reads
\begin{equation} \label{eq:h_decomp}
    H^{\text{Xct}} = T - K^{\text{D}} + 2\delta_S K^{\text{X}},
\end{equation}
where $T$ is a kinetic term related to the propagation of a free electron-hole pair while $-K^{\text{D}}$ is an attractive screened direct interaction and $K^{\text{X}}$ a repulsive bare exchange interaction coupling the electron and hole~\cite{Strinati1988-wy}. $\delta_S$ is 1 for singlets and 0 for triplet excitons.
 
Below we show that the direct interaction is short-range so that  triplet MLXWFs at $\bar{\mathbf{0}}$ and $\bar{\mathbf{R}}$ couple only when the spatial part of their MLXWF's overlap. By contrast the exchange term is long-range and singlet MLXWFs couple even when there is minimal real space overlap. For large $\bar{\mathbf{R}}$, we perform a multipole expansion in this exchange term and show that at leading order MLXWFs are coupled through a dipole-dipole interaction which decays as $1/|\bar{\mathbf{R}}|^3$. In reciprocal space ($\mathbf{Q}$-space), this long-range coupling is non-analytic in the $\mathbf{Q}\rightarrow 0$ limit and can drive a splitting of the longitudinal and transverse branches of the exciton band structure.

\subsection{Exciton Hamiltonian in the MLXWF Basis \label{sec:4A}}

The exciton Hamiltonian is invariant under translations by a  lattice vector, $\bar{\mathbf{R}}$, explicitly 
\begin{equation} \label{eq:h_trans}
H^{\text{Xct}}(\mathbf{R},\mathbf{r}) = H^{\text{Xct}}(\mathbf{R}+\bar{\mathbf{R}},\mathbf{r}),
\end{equation}
so that matrix elements of $H^{\text{Xct}}$ in the Wannier basis obey the following identity   
\begin{equation} \label{eq:h_mat}
\braket{M\bar{\mathbf{R}}|H^{\text{Xct}}|N\bar{\mathbf{R}}'} = \braket{M\bar{\mathbf{0}}|H^{\text{Xct}}|N\bar{\mathbf{R}}'-\bar{\mathbf{R}}}.
\end{equation}
It follows that the matrix element of the exciton Hamiltonian between any two MLXWFs can be constructed from knowledge of 
\begin{equation} \label{eq:h_mat_wann}
    H^{\text{Xct}}_{MN}(\bar{\mathbf{R}}) \equiv \braket{M\bar{\mathbf{0}}|H^{\text{Xct}}|N\bar{\mathbf{R}}}.
\end{equation}

We can relate $H^{\text{Xct}}_{MN}(\bar{\mathbf{R}})$ to the exciton Hamiltonian in the Bloch basis
\begin{equation} \label{eq:h_mat_bloch}
    H^{\text{Xct}}_{SS'}({\mathbf{Q}}) \equiv \braket{S{\mathbf{Q}}|H^{\text{Xct}}|S'{\mathbf{Q}}}.
\end{equation}
Going forward, we use subscripts $``M,N"$ to label matrix elements in the MLXWF basis and $``S,S'"$ to label matrix elements in the Bloch basis. \EDIT{In cases where orbital or band indices are omitted the result is independent of the basis set.}

Using Eq.~\ref{eq:xct_mlwf} and~\ref{eq:xct_mlwf_inv} we find
\begin{equation} \label{eq:h_wann2bloch}
\begin{split}
H_{SS'}^{\text{Xct}}({\mathbf{Q}})
= \sum_{\bar{\mathbf{R}}} e^{i\mathbf{Q}\cdot\bar{\mathbf{R}}} U_{SM}(\mathbf{Q})  H_{MN}^{\text{Xct}}(\bar{\mathbf{R}}) U^{\dagger}_{NS'}(\mathbf{Q}),
\end{split}
\end{equation}
and 
\begin{equation} \label{eq:h_bloch2wann}
\begin{split}
H_{MN}^{\text{Xct}}(\bar{\mathbf{R}}) = \sum_{\mathbf{Q}} e^{-i\mathbf{Q}\cdot\bar{\mathbf{R}}} U_{MS}^\dagger(\mathbf{Q})  H_{SS'}^{\text{Xct}}(\mathbf{Q}) U_{S'N}(\mathbf{Q}),
\end{split}
\end{equation}
where sums over repeated band indices are implied. By construction $H^{\text{Xct}}$ is diagonal in the Bloch basis. Its elements are the exciton eigenenergies, $E_{S\mathbf{Q}}$.

Like $H^{\text{Xct}}$, the direct, $K^{\text{D}}$, and exchange, $K^{\text{X}}$, kernels are translationally invariant so it is sufficient to consider the matrix elements $K^{\text{D}}_{MN}(\bar{\mathbf{R}}) = \braket{{M\bar{\mathbf{0}}}|K^{\rm D}|{N\bar{\mathbf{R}}}}$ and $K^{\text{X}}_{MN}(\bar{\mathbf{R}}) = \braket{{M\bar{\mathbf{0}}}|K^{\rm X}|{N\bar{\mathbf{R}}}}$. 

In the MLXWF basis, the commonly used the static approximation~\cite{Rohlfing2000-wx} to the direct interaction kernel reads
\begin{equation} \label{eq:KD_wann}
\begin{split}
&K^{\text{D}}_{MN}(\bar{\mathbf{R}}) \\
&= \int W_{M\bar{\mathbf{0}}}^\star(\mathbf{R},\mathbf{r}) \varepsilon^{-1}(\mathbf{R},\mathbf{r})\frac{e^2}{r}W_{N\bar{\mathbf{R}}}(\mathbf{R},\mathbf{r}) d\mathbf{R} d\mathbf{r},
\end{split}
\end{equation}
where $e$ is the electronic charge and $\varepsilon^{-1}(\mathbf{R},\mathbf{r})$ is the inverse dielectric function expressed in terms of the average, $\mathbf{R}$, and relative, $\mathbf{r}$, coordinates \EDIT{(see Appendix~\ref{app:C1} for derivation)}. Notably, the direct term involves inter-exciton overlaps of the electron-hole coordinates and only couples $W_{M\bar{\mathbf{0}}}(\mathbf{R},\mathbf{r})$ to $W_{N\bar{\mathbf{R}}}(\mathbf{R},\mathbf{r})$ when there is non-zero overlap in both $\mathbf{R}$ and $\mathbf{r}$. It immediately follows that $K^{\text{D}}_{MN}(\bar{\mathbf{R}})$ decays exponentially in $\bar{\mathbf{R}}$ when the MLXWFs are exponentially localized in $\mathbf{R}$.

Meanwhile the exchange interaction reads
\begin{equation} \label{eq:KX_wann}
\begin{split}
&K^{\text{X}}_{MN}(\bar{\mathbf{R}}) \\
&=  \int W^\star_{M\bar{\mathbf{0}}}(\mathbf{R},\mathbf{r}=\bm{0}) \frac{e^2}{|\mathbf{R}-\mathbf{R}'|} W_{N\bar{\mathbf{R}}}(\mathbf{R}',\mathbf{r}=\bm{0}) d\mathbf{R} d\mathbf{R}'\EDIT{,} 
\end{split}
\end{equation}
\EDIT{(see Appendix~\ref{app:C1} for derivation).}
The exchange interaction involves intra-exciton overlaps of electron-hole pairs -- i.e. the electron and hole within a single exciton must overlap which is why $\mathbf{r}=\mathbf{r}_e-\mathbf{r}_h = \mathbf{0}$. Such overlaps are independent of distance between exciton Wannier functions and for this reason the exchange term couples MLXWFs which reside in cells $\bar{\mathbf{0}}$ and $\bar{\mathbf{R}}$ through the Coulomb interaction even when these MLXWFs are exponentially localized. We expect the exchange interaction to decay slowly with $\bar{\mathbf{R}}$. 

To see how $K^{\text{X}}_{MN}(\bar{\mathbf{R}})$ decays at leading order in $\bar{\mathbf{R}}$ we perform a multipole expansion in the Coulomb interaction. This is justified in the limit where $\bar{\mathbf{R}}$ is much larger than the spatial extent of the MLXWF in the $\mathbf{R}$ coordinate. Retaining only the lowest order dipole-dipole term, we arrive at the dipolar exchange kernel 
\begin{equation} \label{eq:KDip_wann}
\begin{split}
&K^{\text{X,Dip}}_{MN}(\bar{\mathbf{R}}) = e^2 \bigg[ \frac{\mathbf{P}^{\star}_M\cdot \mathbf{P}_{N}}{|\bar{\mathbf{R}}|^3} - 3\frac{(\mathbf{P}^{\star}_M \cdot \bar{\mathbf{R}} )(\mathbf{P}_N \cdot \bar{\mathbf{R}})}{|\bar{\mathbf{R}}|^5} \bigg],
\end{split}
\end{equation}
where $\mathbf{P}_M$ denotes the dipole moment associated with the exciton Wannier function $M$, namely,
\begin{equation} \label{eq:P_wann}
\begin{split}
\mathbf{P}_M =\int \mathbf{R} W_{M\bm{0}} (\mathbf{R},\mathbf{r} = \mathbf{0}) d\mathbf{R},
\end{split}
\end{equation}
\EDIT{(see Appendix ~\ref{app:C2A} for derivation)}.
We see that for $\mathbf{P}_M\ne \mathbf{0}$, $K^{\text{X}}_{MN}(\bar{\mathbf{R}})$ decays like $1/|\bar{\mathbf{R}}|^3$ in the large $\bar{\mathbf{R}}$ limit. 

For later use it is convenient to introduce the exciton dipole moment in the Bloch basis, namely
\begin{equation} \label{eq:P_bloch}
\begin{split}
    \mathbf{P}_{S} =
    \int \mathbf{R} \Psi_{S\mathbf{0}}(\mathbf{R},\mathbf{r}=\mathbf{0}) d\mathbf{R}.
\end{split}
\end{equation}
We emphasize that $\mathbf{P}_S$ is, up to a multiplicative constant, equivalent to the usual exciton transition dipole matrix element used for studying optical processes. The connection is made explicitly in Eq.~\ref{eq:PS}. The $\mathbf{r}=0$ argument, often not explicitly expressed, ensures that $\mathbf{r}_e=\mathbf{r}_h$: the photo-excited state initially creates an electron and hole at the same position so as not to violate causality. 

The exciton dipole matrix elements in the Bloch and Wannier basis are related through the following expressions
\begin{equation} \label{eq:P_bloch2wann}
\begin{split}
    \mathbf{P}_M &= \frac{1}{\sqrt{N_{\mathbf{Q}}}} \sum_S U_{SM}( \mathbf{0}) \mathbf{P}_{S} \\
    \mathbf{P}_{S} &= \sqrt{N_{\mathbf{Q}}}\sum_M U^\dagger_{MS}(\mathbf{0}) \mathbf{P}_M,
\end{split}
\end{equation}
provided $U_{SM}(\mathbf{Q})$ is analytic in the $\mathbf{Q} \rightarrow \bm{0}$ limit \EDIT{(see Appendix~\ref{app:C3} for derivation).}

\subsection{Dipolar Coupling and Non-Analyticity \label{sec:4B}}

In reciprocal space, the dipolar interaction can lead to a non-analyticity in $\mathbf{Q}$ which in turn has important implications for the construction of exponentially localized MLXWFs. To see how this comes about, we Fourier transform $K^{\text{X,Dip}}_{MN}(\bar{\mathbf{R}})$ to find
\begin{equation} \label{eq:KDip_wann_Q}
\begin{split}
K^{\text{X,Dip}}_{MN}&(\mathbf{Q}) = \sum_{\bar{\mathbf{R}}} e^{i\mathbf{Q}\cdot\bar{\mathbf{R}}} K^{\text{X,Dip}}_{MN}(\bar{\mathbf{R}}) \\
&= \frac{4 \pi e^2}{V_{\text{\rm{uc}}}} \sum_{\mathbf{G}} \frac{ [\mathbf{P}^{\star}_M \cdot (\mathbf{Q}+\mathbf{G}) ] [\mathbf{P}^{}_N \cdot (\mathbf{Q}+\mathbf{G}) ]}{|\mathbf{Q}+\mathbf{G}|^2},  \\
\end{split}
\end{equation}
where $\mathbf{G}$ denotes a reciprocal lattice vector, and it is understood that $\mathbf{Q}$ is restricted to the first Brillouin zone \EDIT{(see Appendix~\ref{app:C2B} for derivation)}. Eq.~\ref{eq:KDip_wann_Q} formally diverges for all $\mathbf{Q}$, \EDIT{which can be seen by noting that the summand remains finite for large $\mathbf{G}$}. \EDIT{This divergence is a well known} consequence of not removing the self-interaction term, $K^{\text{X,Dip}}_{MN}(\bar{\mathbf{R}}=\mathbf{0})$, in \EDIT{Eq.~\ref{eq:KDip_wann}}, before Fourier transforming\EDIT{~\cite{Born1956-ty}}. \EDIT{In Sec.~\ref{sec:5} we will regulate this divergence. Presently we are interested in the non-analytic structure of $K_{MN}^{\text{X,Dip}}(\mathbf{Q})$ near the origin, which can be studied at this level despite the divergence. To isolate the small $\mathbf{Q}$ behavior, we define the non-analytic kernel as}
\begin{equation} \label{eq:KNA_def}
K_{MN}^{\text{NA}}(\mathbf{Q}) \equiv \lim_{\mathbf{G}=\mathbf{0},\mathbf{Q}\rightarrow \bm{0}} K_{MN}^{\text{X,Dip}}(\mathbf{Q}),
\end{equation}
\EDIT{where the notation indicates that $\mathbf{G}$ should be set to zero, prior to taking the limit. We then find} 
\begin{equation} \label{eq:KNA_wann_Q}
    K^{\text{NA}}_{MN}(\mathbf{Q}) = \frac{4\pi e^2}{V_{\text{uc}}} \frac{(\mathbf{P}_M^{\star} \cdot\mathbf{Q})(\mathbf{P}^{}_N \cdot\mathbf{Q})}{|\mathbf{Q}|^2}.
\end{equation}
For general $\mathbf{P}_M$, $K^{\text{NA}}_{MN}(\mathbf{Q})$ will approach different values as $\mathbf{Q} \rightarrow \bm{0}$ depending on the path taken in $\mathbf{Q}$-space, which is why we refer to this as \EDIT{the} non-analytic \EDIT{kernel}. The non-analyticity persists in Bloch basis where
\begin{equation} \label{eq:KDip_bloch}
\begin{split}
K^{\text{X,Dip}}_{SS'}(\mathbf{Q}) &= \frac{4 \pi e^2}{V_{\text{uc}}}  \sum_{MN} U_{SM}(\mathbf{Q})  U^{\dagger}_{NS}(\mathbf{Q}) \\
&\times \sum_{\mathbf{G}}\frac{ [\mathbf{P}^{\star}_M \cdot (\mathbf{Q}+\mathbf{G}) ] [\mathbf{P}^{}_N \cdot (\mathbf{Q}+\mathbf{G}) ]}{|\mathbf{Q}+\mathbf{G}|^2}  \\
\end{split}
\end{equation}
and
\begin{equation} \label{eq:KNA_bloch}
    K^{\text{NA}}_{SS'}(\mathbf{Q}) =\frac{4\pi e^2}{V_{\text{uc}}N_{\mathbf{Q}}} \frac{(\mathbf{P}_S^{\star} \cdot\mathbf{Q})(\mathbf{P}^{}_{S'} \cdot\mathbf{Q})}{|\mathbf{Q}|^2}.
\end{equation}
In deriving Eq.~\ref{eq:KNA_bloch} we have made use of Eq.~\ref{eq:P_bloch2wann} to write this expression in terms of $\mathbf{P}_S$. This non-analyticity is well known ~\cite{Heller1951-vk, Hopfield1960-fa, Knox1963-dl, Strinati1988-wy} and was recently revisited in the context of first-principles BSE calculations of exciton dispersion~\cite{Qiu2015-bm, Qiu2021-ll}, where it was calculated by directly taking the $\mathbf{Q}\rightarrow \bm{0}$ limit of $K_{SS'}^{\text{X}}(\mathbf{Q})$. Our present derivation reinterprets this non-analyticity as stemming from dipolar coupling between MLXWFs. 

The non-analytic kernel can lift the degeneracy between longitudinal (L) and transverse (T) exciton states, a phenomena known as LT splitting ~\cite{Heller1951-vk, Hopfield1960-fa, Knox1963-dl, Strinati1988-wy}. Further, in low-symmetry systems, as a direct result of this non-analyticity, singlet exciton eigenvalues can exhibit rapid angular variation about $\mathbf{Q} = \mathbf{0}$, and may appear discontinuous when plotted along certain directions in reciprocal space.

\EDIT{Finally we note that the multipole expansion presented in this section is especially applicable to cases where MLXWFs are centered at the origin of their home cell. In cases with multiple MLXWFs centered at different places in the home cell, it may be more appropriate to expand about the distances between Wannier centers, ie. $\bar{\mathbf{d}}=\bar{\mathbf{R}} +\bm{\tau}_N -\bm{\tau}_M$, where $\bm{\tau}_N$ denotes the position of the Wannier center of the $N^{\text{th}}$ MLXWF in the home cell. The results given in the section are easily generalized in this case but left to future work.}

\subsection{Analogy with Phonons \label{sec:4C}}

The notation, and relation between, $K^{\text{X},\text{Dip}}$ and $K^{\text{NA}}$ resembles that found in the dynamical theory of lattice vibrations where infrared (IR)-active phonons lead to similar dipolar coupling,  LT splittings, and, in low-symmetry systems, a $\mathbf{Q}$-space non-analyticity~\cite{Born1956-ty, Gonze1997-mx}. We summarize the situation for phonons below and then draw an analogy to excitons.

In many materials, polar ionic displacement patterns can induce an electric dipole moment. For long wavelength, IR-active modes, the phonon eigenvector is such that these dipole moments add constructively and we can associate a net microscopic dipole with each unit cell. Coupling between these dipoles drives a splitting of the longitudinal and transverse optical phonon branches and in low-symmetry systems can give rise to a non-analyticity in the dynamical matrix~\cite{Born1956-ty, Gonze1997-mx, Rabe1995-av}.

With the aide of Eq.~\ref{eq:xct_mlwf_inv} we can arrive at a similar physical interpretation for excitons. Notably, Eq.~\ref{eq:xct_mlwf_inv} allows us to express the exciton wavefunction as a sum over MLXWFs. Each MLXWF resides in a cell labeled by $\bar{\mathbf{R}}$, and carries a dipole moment $\mathbf{P}_M$. For bright excitonic states ($\mathbf{P}_S\ne \mathbf{0}$), the Wannier rotation matrices are such that these dipole moments add constructively and we can associate a net microscopic dipole with each unit cell. Long-range interactions between these dipoles drive a splitting of the longitudinal and transverse exciton branches and in low-symmetry systems can give rise to a non-analyticity in the exciton Hamiltonian.

More generally, we note that dipolar interactions, LT-splitting, and non-analyticity are phenomena associated with polar excitations, IR active phonons or bright excitons.

\section{Partitioning the exciton Hamiltonian in the Wannierization subspace \label{sec:5}}

The long-range dipolar coupling in the singlet exciton Hamiltonian is critical to account for in post-processing applications like Wannier-Fourier interpolation, where $\smash{H^{\text{Xct}}_{MN}(\bar{\mathbf{R}})}$ must decay rapidly with increasing $\bar{\mathbf{R}}$ to obtain accurate interpolations. These long-range couplings constitute a non-trivial difference between MLXWFs and their one-electron counterparts. 

A strategy for handling these interactions is to first remove the long-range coupling to arrive at a short-range (SR) Hamiltonian \EDIT{which we expect to be more suitable for use in Wannier-based post-processing applications.}

Accordingly, we partition the exciton Hamiltonian as follows
\begin{equation} \label{eq:H_sr}
    H^{\text{Xct}}(\mathbf{Q})  =  H^{\text{SR}}(\mathbf{Q}) + 2\delta_S K^{\text{LR}}(\mathbf{Q}),
\end{equation}
\EDIT{
where the long-range kernel is defined as 
\begin{equation} \label{eq:Klr}
    K^{\text{LR}}(\mathbf{Q}) = K^{\text{X,Dip}}(\mathbf{Q}) - \bar{K}^{\text{X,Dip}}(\mathbf{0})
\end{equation}
and where expressions for $K^{\text{X,Dip}}(\mathbf{Q})$ are given in the MLXWF and Bloch bases in Eqs.~\ref{eq:KDip_wann_Q} and~\ref{eq:KDip_bloch}, respectively. The bar on $\bar{K}^{\text{X,Dip}}(\mathbf{0})$ indicates that the $\mathbf{G}=\mathbf{0}$ component should be omitted when computing this term. This subtraction regulates the divergence in $K^{\text{X,Dip}}(\mathbf{Q})$ so that $K^{\text{LR}}(\mathbf{Q}\ne \mathbf{0})$ is formally convergent. We have dropped orbital/band indices in these expressions to indicate that the partitioning can be made in any basis set. However, for the sake of concreteness it is convenient to specialize to the MLXWF basis. Then Eq.~\ref{eq:Klr} reads
\begin{equation} \label{eq:K_lr_mn}
\begin{split}
    K^{\text{LR}}_{MN}(\mathbf{Q})
    &=\frac{4 \pi e^2}{V_{\text{\rm{uc}}}} \sum_{\mathbf{G}} \frac{ [\mathbf{P}^{\star}_M \cdot (\mathbf{Q}+\mathbf{G}) ] [\mathbf{P}^{}_N \cdot (\mathbf{Q}+\mathbf{G}) ]}{|\mathbf{Q}+\mathbf{G}|^2} \\
    &-\frac{4 \pi e^2}{V_{\text{\rm{uc}}}} \sum_{\mathbf{G}\ne \mathbf{0}} \frac{ [\mathbf{P}^{\star}_M \cdot \mathbf{G} ] [\mathbf{P}^{}_N \cdot \mathbf{G} ]}{|\mathbf{G}|^2}.
\end{split}
\end{equation}
As written, $K^{\text{LR}}_{MN}(\mathbf{Q})$ is well defined for all $\mathbf{Q}\ne\mathbf{0}$ and approaches the non-analytic kernel in the limit where $\mathbf{Q}\rightarrow \mathbf{0}$. Explicitly 
\begin{equation} \label{eq:KNA_KLR}
    \lim_{\mathbf{Q}\rightarrow\mathbf{0}} K^{\text{LR}}_{MN}(\mathbf{Q}) = K^{\text{NA}}_{MN}(\mathbf{Q}).
\end{equation}
This limit should be compared to that given in  Eq.~\ref{eq:KNA_def}. Note that here there is no need to set $\mathbf{G}=\mathbf{0}$ before the taking the limit. 

At $\mathbf{Q}=\mathbf{0}$ Eq.~\ref{eq:K_lr_mn} evaluates to 
\begin{equation} \label{eq:Klr_Q0}
    K^{\text{LR}}_{MN}(\mathbf{0}) =\frac{4 \pi e^2}{V_{\text{\rm{uc}}}} \frac{ [\mathbf{P}^{\star}_M \cdot \mathbf{G} ] [\mathbf{P}^{}_N \cdot \mathbf{G} ]}{|\mathbf{G}|^2} \bigg|_{\mathbf{G}=\mathbf{0}},
\end{equation}
which is indeterminate. In the following section we show that a similar indeterminate term appears in $H^{\text{Xct}}_{MN}(\mathbf{0})$ which to lowest order cancels with $K^{\text{LR}}_{MN}(\mathbf{0})$. The cancellation is a direct consequence of our partitioning and definition of $K^{\text{LR}}_{MN}(\mathbf{Q})$. The net result is that at lowest order $H_{MN}^{\text{SR}}(\mathbf{0})$ is well defined despite the indeterminate contributions from both $H^{\text{Xct}}_{MN}(\mathbf{0})$ and $K^{\text{LR}}_{MN}(\mathbf{0})$.
}

\EDIT{\subsection{Partitioning for $\mathbf{Q}= \mathbf{0}$\label{sec:5A}} 

The indeterminate behavior of $H^{\text{Xct}}(\mathbf{0})$ originates from the exchange kernel. To see this in detail we Fourier transform the exchange interaction in Eq.~\ref{eq:KX_wann} and set $\mathbf{Q}=\mathbf{0}$ to find
\begin{equation} \label{eq:exchange_g0}
K_{MN}^{\text{X}}(\mathbf{0}) =\sum_\mathbf{G} \braket{M\bar{\mathbf{0}}|e^{i\mathbf{G}\cdot\mathbf{R}}|0}  v(\mathbf{G})  \braket{0|e^{-i\mathbf{G}\cdot\mathbf{R'}}|N\bar{\mathbf{0}}},
\end{equation}
where 
\begin{equation}
   v(\mathbf{G})= \frac{4\pi e^2}{V_{\text{uc}}}\frac{1}{|\mathbf{G}|^2}, 
\end{equation}
and
\begin{equation}
\braket{0|e^{-i\mathbf{G}\cdot\mathbf{R}}|N\bar{\mathbf{0}}}=\int e^{-i\mathbf{G}\cdot\mathbf{R}} W_{N\bar{\mathbf{0}}}(\mathbf{R},\mathbf{0}) d\mathbf{R}.
\end{equation}
Now\EDIT{, we} isolate the head ($\mathbf{G}=\mathbf{0}$) contribution to the sum in Eq.~\ref{eq:exchange_g0} and expand the complex exponential to lowest order to find
\begin{equation} \label{eq:head_KX}
\begin{split}
K_{MN;\mathbf{G}=\mathbf{0}}^{\text{X}}(\mathbf{0})
=  \frac{4\pi e^2}{V_{\text{uc}}} \frac{(\mathbf{P}^\star_M\cdot \mathbf{G})(\mathbf{P}_{N}\cdot \mathbf{G})}{|\mathbf{G}|^2}\bigg|_{\mathbf{G}=\mathbf{0}} + \cdots,
\end{split}
\end{equation}
where we have used $\braket{0|e^{-i\mathbf{G}\cdot\mathbf{R}}|N\bar{\mathbf{0}}} = -i\mathbf{G} \cdot\mathbf{P}_N+\cdots$. 
This limiting behavior should be compared with $K_{MN}^{\text{LR}}(\mathbf{0})$, given in Eq.~\ref{eq:Klr_Q0}. We see that at $\mathbf{Q}=\mathbf{0}$ the long-range kernel exactly equals the lowest order term in the expansion of the head of the exchange kernel. For singlets, while both $H_{MN}^{\text{Xct}}(\mathbf{0})$ and $K_{MN}^{\text{LR}}(\mathbf{0})$ are indeterminate, at lowest order, $H_{MN}^{\text{SR}}(\mathbf{0}) = H_{MN}^{\text{Xct}}(\mathbf{0}) - 2\delta_SK_{MN}^{\text{LR}}(\mathbf{0})$ is well defined; the indeterminate contributions cancel. Indeed it is this cancellation which motivated the definition of $K^{\text{LR}}_{MN}(\mathbf{Q})$ given in Eq.~\ref{eq:Klr}.

When higher order terms in Eq.~\ref{eq:head_KX} are negligible, subtracting $2\delta_SK^{\text{LR}}_{MN}(\mathbf{0})$ from $H_{MN}^{\text{Xct}}(\mathbf{0})$ is equivalent to setting the head of the Coulomb interaction in the exchange term to zero. Explicitly
\begin{equation} \label{eq:Hsr_head}
\begin{split}
    H_{MN}^{\text{SR}}(\mathbf{0}) = T_{MN}(\mathbf{0}) - K^{\text{D}}_{MN}(\mathbf{0}) + 2 \delta_S \bar{K}_{MN}^{\text{X}}(\mathbf{0}),
\end{split}
\end{equation}
where $\bar{K}_{MN}^{\text{X}}(\mathbf{0})$ is equivalent to $K_{MN}^{\text{X}}(\mathbf{0})$ but with the $v(\mathbf{G})$ replaced by the modified Coulomb interaction
\begin{equation} \label{eq:mod_coul}
\bar{v}(\mathbf{G}) = \frac{4\pi e^2}{V_{\text{uc}}} \frac{1}{|\mathbf{G}|^2} (1-\delta_{\mathbf{G}\mathbf{0}}).
\end{equation}

While we have only shown the cancellation between $\smash{K^{\text{X}}_{MN;\mathbf{G}=\mathbf{0}}(\mathbf{0})}$ and $\smash{K^{\text{LR}}_{MN}(\mathbf{0})}$ to first order, it is easy to see that including higher order long-range couplings in $K^{\text{X,LR}}_{MN}(\mathbf{Q})$, e.g. dipole-quadropole, quadropole-quadropole etc., systematically leads to further cancellation between these two terms. Further, the cancellation is independent of the basis set. Motivated by these results, we define the short-range Hamiltonian at $\mathbf{Q}=\mathbf{0}$ to be     
\begin{equation}
H^{\text{SR}}(\mathbf{0}) \equiv T(\mathbf{0}) - K^{\text{D}}(\mathbf{0}) + 2 \delta_S \bar{K}^{\text{X}}(\mathbf{0}), 
\end{equation}
where we have dropped the orbital indices to indicate that the definition is basis set independent. Because $H^{\text{Xct}}(\mathbf{0})$ is ill-defined, it is natural to define the exciton states at $\mathbf{Q}=\mathbf{0}$ as eigenstates of $H^{\text{SR}}(\mathbf{0})$. Notably, our definition of the short-range Hamiltonian coincides with the usual \textit{ab initio} BSE Hamiltonian used for studying optical properties where it can be shown that using the modified Coulomb interaction in Eq.~\ref{eq:mod_coul} facilitates the inclusion of local field effects~\cite{Onida2002-ou}.

In practice, $H^{\text{SR}}(\mathbf{0})$ is constructed and diagonalized 
in a basis of electron-hole product states to obtain exciton eigenstates, $\Psi_{S\mathbf{0}}(\mathbf{R},\mathbf{r})$, and eigenvalues $E_{S\mathbf{0}}$ prior to Wannierization. Throughout this work whenever we reference $\mathbf{Q}=\mathbf{0}$ exciton states, e.g. as in Eq.~\ref{eq:xct_mlwf} or~\ref{eq:P_bloch}, we have in mind the eigenstates of $H^{\text{SR}}(\mathbf{0})$.

}

\EDIT{
\subsection{Partitioning for $\mathbf{Q}\ne \mathbf{0}$ \label{sec:5B}}

Working in a basis of $\mathbf{Q}=\mathbf{0}$ exciton states, we can construct an exciton Hamiltonian for states near $\mathbf{Q}=\mathbf{0}$ as follows
\begin{equation} \label{eq:HQ_near0}
\lim_{\mathbf{Q}\rightarrow \mathbf{0}} H_{SS'}^{\text{Xct}}(\mathbf{Q}) = H_{SS'}^{\text{SR}}(\mathbf{0}) + 2\delta_S K_{SS'}^{\text{NA}}(\mathbf{Q})
\end{equation}
where $H_{SS'}^{\text{SR}}(\mathbf{0})=E_{S\mathbf{0}}\delta_{SS'}$ and $K_{SS'}^{\text{NA}}(\mathbf{Q})$ was defined in Eq.~\ref{eq:KNA_bloch}. 

When we try to rotate Eq.~\ref{eq:HQ_near0} into the MLXWF basis using the exciton rotation matrices, $U_{MS}(\mathbf{Q})$, we encounter an challenge. Because Wannierization is always performed in a subspace, it is only possible to rotate the sector of $\lim_{\mathbf{Q}\rightarrow \mathbf{0}}H^{\text{Xct}}_{SS'}(\mathbf{Q})$ for which $S,S'\in\mathcal{W}$. By contrast the non-analytic interaction mixes the entire space of bright ($\mathbf{P}_S\ne 0$) excitonic states. Inevitably, the long-range interaction couples MLXWFs with states outside the Wannierization subspace. 

At lowest order we could simply neglect the coupling to states outside of the Wannierization window, however numerical results indicate that this tends to greatly exaggerate the effect of $K_{SS'}^{\text{NA}}(\mathbf{Q})$, leading to nonphysical results. Our approach here is to first derive an effective Hamiltonian which downfolds the non-analytic interaction into the Wannierization subspace and then rotate the Hamiltonian into the MLXWF basis.

Let $S,S'\in \mathcal{W}$ denote excitonic states in the Wannierization window (also referred to as the active space) and $R,R'\notin \mathcal{W}$ label excitonic states outside the Wannierization window (the passive space). Adopting a Lowdin partitioning strategy~\cite{Lowdin1951-tq}, we treat $H_{SS'}^{\text{SR}}(\mathbf{0})$ as our unperturbed Hamiltonian, $K_{SS'}^{\text{NA}}(\mathbf{Q})$ as a perturbation, and sum over all transitions involving states outside the Wannierization subspace, effectively integrating out the passive states. We arrive at the following expression
\begin{widetext}
\begin{equation} \label{eq:RSPT}
\lim_{\mathbf{Q}\rightarrow \mathbf{0}}H^{\text{Xct,eff}}_{SS'}(\mathbf{Q},\omega) = H^{\text{SR}}_{SS'}(\mathbf{0}) + 2\delta_S\bigg[ K^{\text{NA}}_{SS'}(\mathbf{Q}) + 2 \sum_{R\notin \mathcal{W}}\frac{K^{\text{NA}}_{SR}(\mathbf{Q})K^{\text{NA}}_{RS'}(\mathbf{Q})}{\omega-E_{R\mathbf{0}}} +  2^2\sum_{R,R'\notin \mathcal{W}}\frac{K^{\text{NA}}_{SR}(\mathbf{Q})K^{\text{NA}}_{RR'}(\mathbf{Q})K^{\text{NA}}_{R'S'}(\mathbf{Q})}{(\omega-E_{R\mathbf{0}})(\omega-E_{R'\mathbf{0}})} + \cdots \bigg],
\end{equation}
\end{widetext}
where the superscript ``eff" serves as a reminder that this effective Hamiltonian mixes only states in the Wannierization subspace. Our analysis implies that even after downfolding, the exciton Hamiltonian can still be partitioned into a short and long-range part 
\begin{equation} \label{eq:H_downfolded}
\lim_{\mathbf{Q}\rightarrow \mathbf{0}} H^{\text{Xct,eff}}_{SS'}(\mathbf{Q},\omega) = H^{\text{SR}}_{SS'}(\mathbf{0}) + 2\delta_S K_{SS'}^{\text{NA,eff}}(\mathbf{Q},\omega),
\end{equation}
where $K_{SS'}^{\text{NA,eff}}(\mathbf{Q},\omega)$ denotes the bracketed term in Eq.~\ref{eq:RSPT} summed to infinite order. The price to pay for downfolding is that the non-analytic part is now frequency dependent and Eq.~\ref{eq:H_downfolded} must be solved self-consistently to obtain exciton eigenvalues. When all terms in the infinite series in Eq.~\ref{eq:RSPT} are retained, eigenvalues of this effective Hamiltonian exactly coincide with those of the original Hamiltonian in Eq.~\ref{eq:HQ_near0} for $S\in \mathcal{W}$.

Fortunately the separable structure of $K^{\text{NA}}_{SS'}(\mathbf{Q})$ allows the series to be summed to infinite order to obtain the exact result 
\begin{equation} \label{eq:KNA_eff}
\begin{split}
    K^{\text{NA,eff}}_{SS'}(\mathbf{Q},\omega)
    = \varepsilon_{\text{BG}}^{-1}(\mathbf{Q},\omega) K_{SS'}^{\text{NA}}(\mathbf{Q}),
\end{split}
\end{equation}
where $S,S'\in\mathcal{W}$ and $\bm{\varepsilon}_{\text{BG}}(\mathbf{Q},\omega)$ is the ``background" dielectric function and is given by
\begin{equation} \label{eq:dielectric}
    \varepsilon_{\text{BG}}(\mathbf{Q},\omega) = 1 - \frac{8\pi e^2}{V_{\text{uc}}N_{\mathbf{Q}}|\mathbf{Q}|^2} \sum_{R \notin \mathcal{W}} \frac{(\mathbf{P}_{R}\cdot\mathbf{Q}) ( \mathbf{P}^{\star}_{R} \cdot \mathbf{Q})}{\omega - E_{R\mathbf{0}}},
\end{equation}
where $\mathbf{P}_R$ are dipole moments associated with Bloch exciton (see Eq.~\ref{eq:P_bloch}). Importantly the sum on $R$ in $\varepsilon_{\text{BG}}(\mathbf{Q},\omega)$ runs only over Bloch exciton states outside the Wannierization subspace (see Appendix~\ref{app:D} for derivation).
}

We see the effect of downfolding is for states in the passive space to screen the non-analytic coupling between excitons in the active space. The result is closely related to the $S$ approximation, first introduced in Ref.~\cite{Benedict2002-bt, Qiu2021-po}, where the exchange term in the BSE is screened to compensate for the fact that \EDIT{the BSE} is solved in a finite subspace of electron-hole transitions. Similar downfolding techniques were recently used to study x-ray absorption spectra in liquid water~\cite{Tang2022-sh}. 

\EDIT{
Using Wannier rotation matrices at $\mathbf{Q}=\mathbf{0}$, we can rotate Eq.~\ref{eq:H_downfolded} into the Wannier basis to find
\begin{equation} \label{eq:H_downfolded_MN}
\lim_{\mathbf{Q}\rightarrow \mathbf{0}} H^{\text{Xct,eff}}_{MN}(\mathbf{Q},\omega) = H^{\text{SR}}_{MN}(\mathbf{0}) + 2\delta_S K_{MN}^{\text{NA,eff}}(\mathbf{Q},\omega).
\end{equation}
Let $E_{S\mathbf{Q}}$ and $C_{MS}(\mathbf{Q})$ denote the eigenvalues and eigenvectors of Eq.~\ref{eq:H_downfolded_MN}, respectively. By construction $E_{S\mathbf{Q}}$ exactly coincide with the eigenenergies obtain by diagonalizing Eq.~\ref{eq:HQ_near0} for $S\in \mathcal{W}$. 

Once $E_{S\mathbf{Q}}$ and $C_{MS}(\mathbf{Q})$ are known, it is possible to cast Eq.~\ref{eq:H_downfolded_MN} as a static, albeit non-Hermitian, Hamiltonian, namely
\begin{equation} \label{eq:H_downfolded_static}
    \lim_{\mathbf{Q}\rightarrow \mathbf{0}} H^{\text{Xct,eff}}_{MN}(\mathbf{Q}) = H^{\text{SR}}_{MN}(\mathbf{0}) + 2\delta_S K_{MN}^{\text{NA,eff}}(\mathbf{Q}),
\end{equation}
where
\begin{equation}
    K^{\text{NA},\text{eff}}_{MN}(\mathbf{Q}) = \sum_{N'}  K^{\text{NA}}_{MN'}(\mathbf{\mathbf{Q}}) [\varepsilon^{-1}_{\text{BG}}(\mathbf{Q})]_{N'N}
\end{equation}
with
\begin{equation} \label{eq:epsilon_MN}
    [\varepsilon^{-1}_{\text{BG}}(\mathbf{Q})]_{N'N} = \sum_{S} C_{N'S}(\mathbf{Q}) \varepsilon^{-1}_{\text{BG}}(\mathbf{Q},E_{S\mathbf{Q}})  C^{-1}_{SN}(\mathbf{Q})
\end{equation}
and with $K^{\text{NA}}_{MN}(\mathbf{\mathbf{Q}})$ given explicitly in Eq.~\ref{eq:KNA_bloch}. Derivations of these results are given in Appendix~\ref{app:D}, where we further show that Eq.~\ref{eq:H_downfolded_static}, has the same eigenvalues and eigenvectors as Eq.~\ref{eq:H_downfolded_MN}. 
}

\EDIT{
To recap, we have developed a partitioning of the exciton Hamiltonian near $\mathbf{Q}=\mathbf{0}$ which (1) mixes only states in the Wannierization subspace, (2) reproduces the exciton eigenvalues near $\mathbf{Q}=\mathbf{0}$, and (3) decouples into a short-range and long-range (non-analytic) part.

Motivated by these results, we define an effective short-range Hamiltonian near $\mathbf{Q}=\mathbf{0}$ as
\begin{equation} \label{eq:part_small_Q}
    \lim_{\mathbf{Q}\rightarrow \mathbf{0}} H_{MN}^{\text{SR,eff}}(\mathbf{Q}) = \lim_{\mathbf{Q}\rightarrow \mathbf{0}} H^{\text{Xct}}_{MN}(\mathbf{Q}) -2 \delta_S K^{\text{NA},\text{eff}}_{MN}(\mathbf{Q}),
\end{equation}
where $H^{\text{Xct}}_{MN}(\mathbf{Q})$ is the original exciton Hamiltonian. We can generalize this definition to arbitrary $\mathbf{Q}$ as follows
\begin{equation} \label{eq:part_Q}
    H_{MN}^{\text{SR},\text{eff}}(\mathbf{Q}) =  H^{\text{Xct}}_{MN}(\mathbf{Q}) -2 \delta_S K^{\text{LR},\text{eff}}_{MN}(\mathbf{Q}),
\end{equation}
where
\begin{equation} \label{eq:k_lr_eff}
    K_{MN}^{\text{LR},\text{eff}}(\mathbf{Q}) = \sum_{N'} K^{\text{LR}}_{MN'}(\mathbf{\mathbf{Q}}) [\varepsilon^{-1}_{\text{BG}}(\mathbf{Q})]_{N'N}
\end{equation}
and with $[\varepsilon^{-1}_{\text{BG}}(\mathbf{Q})]_{N'N}$ and $K^{\text{LR}}_{MN}(\mathbf{Q})$ defined in Eqs.~\ref{eq:epsilon_MN} and~\ref{eq:K_lr_mn}, respectively. From Eq.~\ref{eq:KNA_KLR} it immediately follows that when $\mathbf{Q}\rightarrow \mathbf{0}$ Eq.~\ref{eq:part_Q} reduces to Eq.~\ref{eq:part_small_Q}. 

In practice, $K^{\text{LR,eff}}_{MN}(\mathbf{Q})$ is constructed as follows. First, we build and diagonalizing Eq.~\ref{eq:H_downfolded_MN} to obtain $E_{S\mathbf{Q}}$ and $C_{MS}(\mathbf{Q})$.  With these eigenvalues and eigenvectors, we construct $[\varepsilon^{-1}_{\text{BG}}(\mathbf{Q})]_{N'N}$ according to Eq.~\ref{eq:epsilon_MN}. The effective long-range kernel is then constructed according to Eqs.~\ref{eq:k_lr_eff}. The procedure is computationally inexpensive, involving only a few digiagonalizations of a small $N_{\mathcal{W}}\times N_{\mathcal{W}}$ matrix at each $\mathbf{Q}$-point.

}

We emphasize that all results presented here are for bulk materials where the $\mathbf{Q}\rightarrow \mathbf{0}$ limit of the Coulomb interaction is given by $v(\mathbf{Q})=4 \pi e^2 / |\mathbf{Q}|^2$. For confined and low dimensional systems, the $\mathbf{Q} \rightarrow \mathbf{0}$ limit of $v(\mathbf{Q})$ changes~\cite{Ismail-Beigi2006-zo} and  and expressions in this section should be modified accordingly. 

\EDIT{We have further neglected the effects of spin-orbit coupling. When included, the exciton Hamiltonian can no longer be partitioned into a singlet and triplet sector, instead excitons should be labeled by their total angular momentum. In general all excitons receive a exchange contribution and the partitioning described in this section should be applied to all states. In practice, the formulae of this section can be used as they stand provided the definition of $\mathbf{P}_S$ in Eq.~\ref{eq:P_bloch} is appropriately generalized for spinors.}

\section{Wannier-Fourier interpolation \label{sec:6}}

We now turn to the interpolation of exciton band structures, making use of the results derived in the previous section. We start with triplet excitons, where there is no exchange coupling and hence no dipolar interactions which simplifies matters. Here, the interpolation procedure is entirely analogous to what is done at the one-electron level~\cite{Marzari2012-qr, Giustino2007-kc}. We then discuss what modifications must be made to carry out the procedure for singlets.

\subsection{Triplets}

Given the exciton energies, $E_{S\mathbf{Q}_c}$, from, e.g. an \textit{ab initio} BSE calculation, and exciton Wannier rotation matrices, $U_{MS}(\mathbf{Q}_c)$, on a coarse uniform grid of $\mathbf{Q}_c$-points, we first Wannier-Fourier transform to obtain hopping matrix elements
\begin{equation}
H^{\text{Xct}}_{MN}(\bar{\mathbf{R}})= \sum_{S\mathbf{Q}_c} e^{-i\mathbf{Q}_c\cdot\bar{\mathbf{R}}} U^{\dagger}_{MS}(\mathbf{Q}_c) E_{S\mathbf{Q}_c} U_{SN}(\mathbf{Q}_c),
\end{equation}
and subsequently Fourier transform to an arbitrary $\mathbf{Q}$-point to find
\begin{equation}
    H^{\text{Xct}}_{MN}(\mathbf{Q}) = \sum_{\bar{\mathbf{R}}} e^{i\mathbf{Q}\cdot\bar{\mathbf{R}}} H^{\text{Xct}}_{MN}(\bar{\mathbf{R}}).
\end{equation}
Triplet exciton eigenenergies at $\mathbf{Q}$ are obtained by diagonalizing $H^{\text{Xct}}_{MN}(\mathbf{Q})$. The success of this scheme hinges critically on the rapid decay of $H^{\text{Xct}}_{MN}(\bar{\mathbf{R}})$ with increasing $\bar{\mathbf{R}}$, which, for triplets, is guaranteed when $W_{M\bar{\mathbf{0}}}(\mathbf{R},\mathbf{r})$ is well localized in $\mathbf{R}$. 

\subsection{Singlets}

For singlet excitons, even when $W_{M\bar{\mathbf{0}}}(\mathbf{R},\mathbf{r})$ is well-localized in $\mathbf{R}$, the matrix element $H^{\text{Xct}}_{MN}(\bar{\mathbf{R}})$ will still decay as $1/|\bar{\mathbf{R}}|^3$ as discussed in Sec.~\ref{sec:4A}, which in many cases will be too gradual to perform Wannier-Fourier interpolation efficiently. 

To deal with this issue, we remove the long-range interaction before Wannier-Fourier interpolating, and add it back after the interpolation is complete. This strategy is inspired by a widely used technique for Fourier interpolating phonon eigenfrequencies in many semiconductors and insulators~\cite{Gonze1994-gw}. We outline our four-step procedure below.

\begin{enumerate}[leftmargin=*]
\item Subtract the \EDIT{effective long-range} interaction,  \EDIT{defined in Eq.~\ref{eq:k_lr_eff}} to obtain \EDIT{the effective} short range Hamiltonian on a coarse grid:
\begin{equation}
H^{\text{SR}\EDIT{\text{,eff}}}_{MN}(\mathbf{Q}_c) =
H_{MN}^{\text{Xct}}(\mathbf{Q}_c)-2\delta_S K_{MN}^{\text{LR},\EDIT{\text{eff}}}(\mathbf{Q}_c). 
\end{equation}
\item Fourier interpolate to obtain $H^{\text{SR,\EDIT{eff}}}$ at an arbitrary $\mathbf{Q}$-point:
\begin{equation}
\begin{split}
H^{\text{SR,\EDIT{eff}}}_{MN}(\bar{\mathbf{R}})&= \sum_{\mathbf{Q}_c} e^{-i\mathbf{Q}_c \cdot \bar{\mathbf{R}}} H_{MN}^{\text{SR,\EDIT{eff}}}(\mathbf{Q}_c), \\
H_{MN}^{\text{SR,\EDIT{eff}}}(\mathbf{Q}) &= \sum_{\bar{\mathbf{R}}} e^{i\mathbf{Q}\cdot \bar{\mathbf{R}}} H_{MN}^{\text{SR,\EDIT{eff}}}(\bar{\mathbf{R}}).
\end{split}
\end{equation}
\item Add back the effective \EDIT{long-range} interaction to obtain $H^{\text{Xct}}$ at finite $\mathbf{Q}$:
\begin{equation}
    H^{\text{Xct}}_{MN}(\mathbf{Q}) = H_{MN}^{\text{SR\EDIT{,\text{eff}}}}(\mathbf{Q}) + 2\delta_S K_{MN}^{\text{LR},\EDIT{\text{eff}}}(\mathbf{Q}).
\end{equation}
\item Diagonalize $H_{MN}^{\text{Xct}}(\mathbf{Q})$ to obtain singlet exciton eigenvalues at arbitrary $\mathbf{Q}$-points.
\end{enumerate} 

We emphasize here that it is critical to use the effective long-range interaction, defined in Eq.~\ref{eq:k_lr_eff}, rather than the bare long-range interaction, defined in Eq.~\ref{eq:K_lr_mn}. The former properly accounts for long-range coupling to states outside the Wanneirization window while the latter does not.

\section{Implementation and Computational Details \label{sec:7}} 
As previously emphasized, the Wannierization procedure is independent of computational method used to compute excitonic properties. Popular methods include time-dependent density functional theory~\cite{Onida2002-ou}, equation-of-motion coupled cluster~\cite{Wang2020-jq}, and the \textit{ab initio} $GW$-Bethe-Salpeter equation (BSE) approach~\cite{Rohlfing2000-wx} to name a few. Here we adopt the latter method and summarize the approach below.

\subsection{Implementation}

For computational tractability, the BSE is cast as an eigenvalue equation in a basis of electron-hole product states $\braket{\mathbf{r}_e\mathbf{r}_h|cv\mathbf{k}\mathbf{Q}} \equiv \psi_{c \mathbf{k}}(\mathbf{r}_e) \psi^\star_{v \mathbf{k}-\mathbf{Q}}(\mathbf{r}_h)$ and reads
\begin{equation} \label{eq:BSE2}
\begin{split}
(&\varepsilon_{c\mathbf{k}} - \varepsilon_{v\mathbf{k}-\mathbf{Q}})A_{cv\mathbf{k}}^{S\mathbf{Q}} \\
&+ \sum_{c'v'\mathbf{k}'} \braket{cv\mathbf{k}\mathbf{Q}|K^{\text{eh}}|c'v'\mathbf{k}'\mathbf{Q}} A^{S\mathbf{Q}}_{c'v'\mathbf{k}'}
=E_{S\mathbf{Q}}A^{S\mathbf{Q}}_{cv\mathbf{k}},
\end{split}
\end{equation}
where $\varepsilon_{n\mathbf{k}}$ are $GW$ quasiparticle energies and $K^{\text{eh}}=-K^{\text{D}}+2\delta_S K^{\text{X}}$ is the electron-hole interaction kernel. Upon solving the BSE, one obtains exciton eigenenergies, $E_{S\mathbf{Q}}$, and exciton wavefunctions in terms of the expansion coefficients, $A^{S\mathbf{Q}}_{cv\mathbf{k}}$, see Eq.~\ref{eq:bse_exciton}. 

At $\mathbf{Q}=\mathbf{0}$, the exchange term is ill-defined because the Coulomb interaction is divergent in the long-wavelength limit. \EDIT{As discussed in Sec.~\ref{sec:5A}} we build $K^{\text{X}}(\mathbf{Q}=\mathbf{0})$ with the modified Coulomb interaction, $\bar{v}(\mathbf{G}) = 4\pi e^2 /|\mathbf{G}|^2 (1 - \delta_{\mathbf{G}\mathbf{0}})$, which is equivalent to the usual Coulomb interaction except the $\mathbf{G}=\mathbf{0}$ component has been set to zero. For all other $\mathbf{Q}$, the usual Coulomb interaction, $v(\mathbf{Q}+\mathbf{G}) = 4\pi e^2 / |\mathbf{Q}+\mathbf{G}|^2$, is used to construct $K^{\text{X}}(\mathbf{Q})$. Modifying the Coulomb interaction in this way regulates the non-analyticity contained in $K^{\text{X}}(\mathbf{Q})$ and ensures that the exciton coefficients are well defined at $\mathbf{Q}=\mathbf{0}$. While this is not the only regulation scheme, it is convenient and \EDIT{consistent with our definition of the long-range kernel in Eq.~\ref{eq:Klr}}. 

The exciton dipole matrix elements, in the Bloch basis are computed from exciton expansion coefficients and single-particle Bloch states as follows
\begin{equation} \label{eq:PS}
\begin{split}
   \mathbf{P}_{S} &= 
   \int \mathbf{R} \Psi_{S\mathbf{0}}(\mathbf{R},\mathbf{r}=\mathbf{0}) d\mathbf{R} \\
   &= \sqrt{\frac{N_{\mathbf{Q}}}{N_{\mathbf{k}}}}  \sum_{cv\mathbf{k}} A^{S\mathbf{0}}_{cv\mathbf{k}} \braket{u_{v\mathbf{k}}|\mathbf{R}|u_{c\mathbf{k}}}_{\text{uc}}, 
\end{split}
\end{equation}
where $A^{S\mathbf{0}}_{cv\mathbf{k}}$ should be calculated from the modified $\mathbf{Q}=\mathbf{0}$ kernel as discussed in the previous paragraph. Note $\mathbf{P}_S$ is equivalent to the usual exciton transition dipole matrix elements used in optical studies~\cite{Rohlfing2000-wx}. From Eq.~\ref{eq:P_bloch2wann} and Eq.~\ref{eq:PS} it immediately follows that
\begin{equation} \label{eq:PM_practical}
    \mathbf{P}_M = \frac{1}{\sqrt{N_{\mathbf{k}}}} \sum_{S} U_{SM}(\mathbf{0}) \sum_{cv\mathbf{k}} A^{S\mathbf{Q}}_{cv\mathbf{k}} \braket{u_{v\mathbf{k}}|\mathbf{R}|u_{c\mathbf{k}}}_{\text{uc}}.
\end{equation}

\begin{figure}
    \centering
    \includegraphics[scale=0.42]{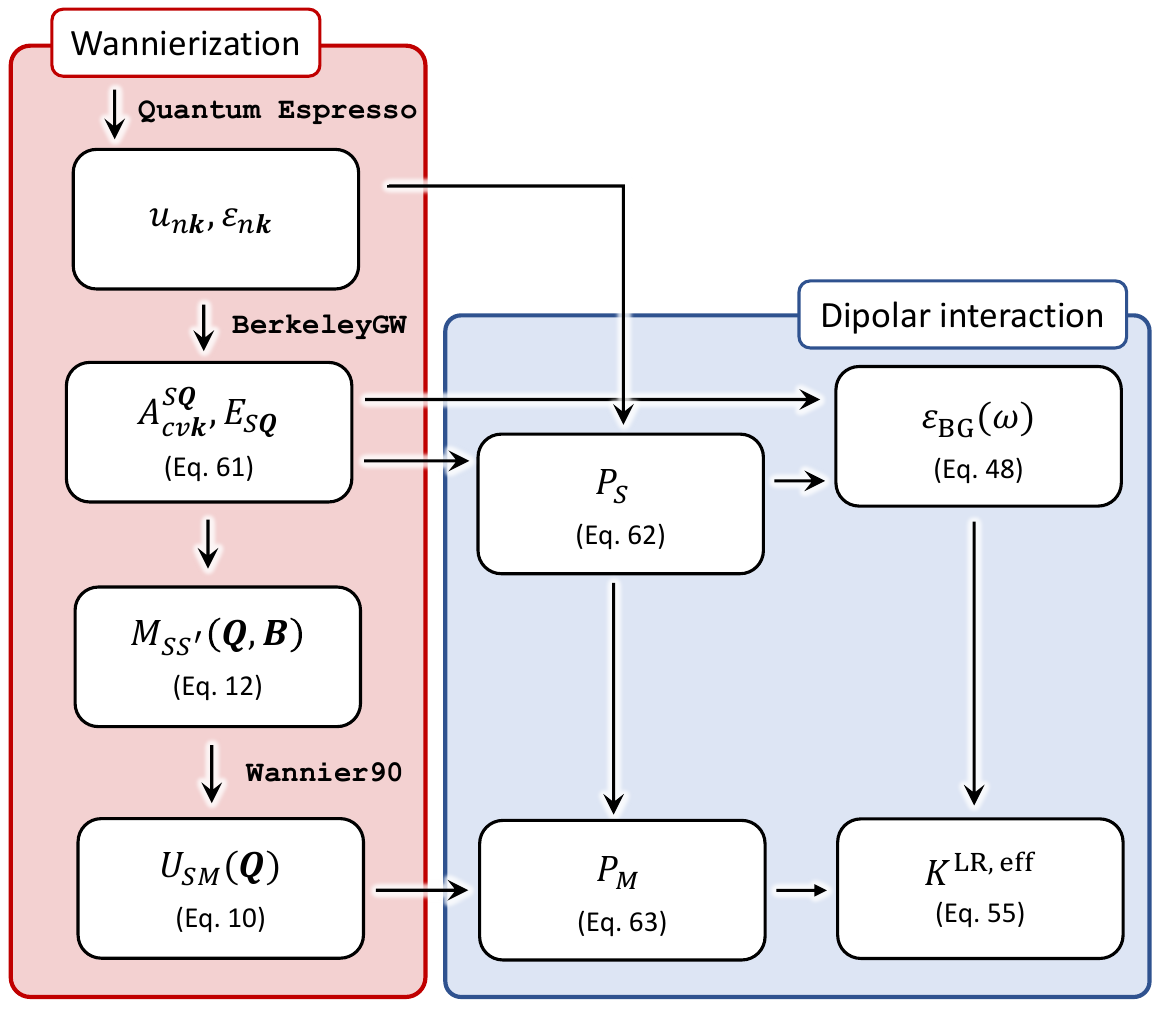}
    \caption{Workflows for obtaining the unitary matrix $U_{SM}(\mathbf{Q})$ that leads to MLXWFs (red section) and computing the dipolar coupling between MLXWFs (blue section). Software used in this work is indicated.}
    \label{fig:flowchart}
\end{figure}

\EDIT{With $\mathbf{P}_M$, we construct the $K_{MN}^{\text{LR},\text{eff}}(\mathbf{Q},\omega)$ as prescribed in Eq.~\ref{eq:k_lr_eff} and detailed in the bottom right-hand corner of Fig.~\ref{fig:flowchart}. In practice Ewald's technique is used to efficiently evaluate Eq.~\ref{eq:k_lr_eff}. Additional details can be found in Appendix~\ref{app:E}.
}

\subsection{Computational Details}

As a first application, we apply the above formalism to Wannierize the low-lying excitons in LiF, a prototypical wide band-gap insulator with weak dielectric screening and correspondingly strong electron-hole interactions. We note that LiF was one of the first systems studied within the \textit{ab initio} $GW$ plus BSE approach~\cite{Rohlfing2000-wx} while the closely related compound LiCl appeared in the original work on MLWFs~\cite{Marzari1997-kz}.

The starting point for all calculations is a ground state density functional theory (DFT) calculation to obtain Kohn-Sham energies and eigenstates~\cite{Hohenberg1964-ku,Kohn1965-zd}. We use a planewave basis set, the local density approximation (LDA), and norm-conserving pseudopotentials taken from \texttt{pseudo-dojo}~\cite{Hamann2013-df,Van_Setten2018-lg}. To converge the ground state density we use an 80 Ry planewave cutoff and $8\times 8 \times 8$ $\mathbf{k}$-mesh. All DFT calculations are performed with \texttt{Quantum Espresso}~\cite{Giannozzi2017-qu}.

We perform BSE calculations for LiF atop the Kohn-Sham eigenvalues and eigenstates. The electron-hole kernel, $K^{\text{eh}}$, is expanded in 1 conduction band and 3 valence bands. For the direct kernel we assume static screening, and compute the static dielectric matrix within the random phase approximation using a sum-over-bands approach~\cite{Adler1962-gy,Wiser1963-qf}. In computing the susceptibility we include 195 unoccupied bands and use a 20 Ry planewave cutoff. All BSE calculations are performed with \texttt{BerkeleyGW}~\cite{Deslippe2012-bq}.

Our workflow requires exciton eigenenergies, $E_{S\mathbf{Q}}$, and eigenstates, $\Psi_{S\mathbf{Q}}$, on a regular $\mathbf{Q}$-mesh. In principle the $\mathbf{k}$- and $\mathbf{Q}$-meshes on which we solve the BSE are decoupled, however in practice it is convenient to take them to be commensurate so that a single set of wavefunctions can be used to compute overlaps appearing in Eq.~\ref{eq:M_SS}. From Eq.~\ref{eq:M_SS} we see that for a fixed $\mathbf{Q}$-mesh we require the following exciton expansion coefficients $A^{S\mathbf{Q}}_{cv\mathbf{k}+\mathbf{B}/2}$ and wavefunction overlaps $\braket{u_{v'\mathbf{k}-\mathbf{Q}-\mathbf{B}/2}|u_{v\mathbf{k}-\mathbf{Q}}}_{\text{uc}}$. Because of the $\mathbf{B}/2$ shifts, for a fixed $\mathbf{Q}$-mesh we require a $\mathbf{k}$-mesh which is both commensurate and twice as dense, along each reciprocal lattice vector direction, as the $\mathbf{Q}$-mesh. To give an explicit example, Wannierizing the exciton on a $5 \times 5 \times 5$ $\mathbf{Q}$-mesh requires solving the BSE at 125 $\mathbf{Q}$-pts with the electron-hole kernel constructed on a $10\times10\times10$ $\mathbf{k}$-mesh. Throughout this work, we use a $\Gamma$ centered $\mathbf{Q}$-mesh and half-shifted $\mathbf{k}$-mesh.

The factor $\mathbf{B}/2$ is linked to our choice to Wannierize in $\mathbf{R}=(\mathbf{r}_e + \mathbf{r}_h)/2$. As discussed Appendix~\ref{app:B} it is more generally possible to Wannierize in $\mathbf{R}=\alpha \mathbf{r}_e + \beta \mathbf{r}_h$, where $\alpha+\beta = 1$ and $\alpha,\beta >0$. Notably for $\alpha=1,\beta=0$ or $\alpha=0,\beta=1$, it is possible to take the two meshes perfectly commensurate. For illustrative purposes, in the main text we have taken $\alpha=1/2,\beta=1/2$ which represents a good compromise between conceptual simplicity and computational tractability. \EDIT{We have numerically confirmed that for LiF the choices $\alpha=0,\beta=1$ and $\alpha=1,\beta=0$ give MLXWFs with the same orbital shapes and centers as those discussed below.}

Finally we note that the total number of exciton bands is fixed by the basis size, in this case $N_{\text{Xct}} = N_{\mathbf{k}} \times N_c \times N_v$, where $N_{c}$ and $N_{v}$ are the number of conduction and valence states used to expand the electron-hole kernel. For the example above we have $N_{\text{Xct}}=3000$. In this work, no symmetry or interpolation scheme is used when building the electron-hole kernel.

\section{Results \label{sec:8}}

\subsection{Exciton Energies and Dispersion}

LiF takes up a rock-salt crystal structure and we use a lattice constant of 4.026\AA\ for all calculations. It is a direct gap material with a LDA Kohn-Sham gap of $E^{\text{LDA}}_g =8.87$ eV, consistent with previously reported values between $8.3-8.8$ eV for the LDA gap~\cite{Shirley1998-xb,Shirley1996-yo,Wang2003-bh,Deilmann2018-px}. We attribute this range to slight variability in the lattice parameters, the inclusion or non-inclusion of the 1s state in the Li pseudopotential, and a mixture of Gaussian vs. planewave basis sets used. To account for quasiparticle corrections we rigidly shift the conduction bands so that the fundamental gap is $E^{\text{QP}}_{\text{gap}} = 14.3$ eV and stretch the valence bands by 15\%, as was done in prior work~\cite{ Benedict1998-yn, Gatti2013-rb}.

Our BSE calculation of LiF shows that the lowest energy triplet ($T$) exciton, at $\Gamma$ $(\mathbf{Q}=\mathbf{0})$, is three-fold degenerate. The additional repulsive exchange interaction in the singlet kernel lifts the degeneracy, leading to a two-fold degenerate transverse ($S_t$), and non-degenerate longitudinal ($S_l$) exciton, as expected. We obtain the following energies for these states: $E_T=12.00$ eV, $E_{S_t}=12.36$ eV and $E_{S_l}=13.08$ eV, all of which are largely consistent with prior work. Our exciton binding energy, $E_{\text{B}} = E^{\text{QP}}_{\text{gap}} - E_{S_t} = 1.94$ eV, is slightly larger than what has been reported in the older literature ($1.5-1.8$ eV)~\cite{Rohlfing2000-wx, Benedict1998-yn, Wang2003-bh} and slightly less than a recent report of a $2.05$ eV~\cite{Sun2020-gt}. Similarly, the LT-splitting $\Delta_{\text{LT}} = E_{S_l}-E_{S_t}=0.72$ eV is somewhat larger than $0.5$ eV previously reported~\cite{Rohlfing2000-wx}. For present purposes our results are in satisfactory agreement with prior work.  

In Fig.~\ref{fig:exciton_dispersion} we show the BSE-computed exciton dispersion (blue curve) for the low energy triplet (right panel) bands. We find that the excitons at $\Gamma$ are three-fold degenerate. The three lowest exciton bands remain entangled throughout the Brillouin zone but are well separated from higher lying bands (see Fig.~\ref{fig:exciton_dispersion}). In the left panel of the same figure, we show the exciton dispersion for the singlet states. We can clearly see that the three-fold degeneracy at $\Gamma$ is lifted by the exchange interaction, with the longitudinal branch higher in energy than the two degenerate transverse branches~\cite{Onodera1967-pj}. The eigenvalues approach a well defined value in the $\mathbf{Q}\rightarrow 0$ limit, independent of the path taken the $\mathbf{Q}$-space. The three lowest singlet bands are not entirely isolated from higher lying singlet states, however the disentanglement is fairly minimal occurring only near the L points at the BZ edge. Our calculations are in very good agreement with prior \textit{ab initio} calculations of the exciton dispersion in LiF along high-symmetry paths~\cite{Gatti2013-rb}.

\subsection{Wannierization of Excitons in LiF}

Based on the exciton dispersion, we restrict our analysis to the subspace $\EDIT{\mathcal{W}}$ which contains the lowest three exciton bands, $N_{\EDIT{\mathcal{W}}}=3$. For triplet states we do not provide a starting guess for the Wannierization procedure. For singlets, which are minimally entangled, we find that it is helpful to provide \texttt{Wannier90} with some initial guess for the exciton Wannier functions. Here we use the use the already Wannierized triplet states as our initial guess. In Appendix~\ref{app:B} we detail how these overlaps are constructed. We utilize the standard disentanglement scheme~\cite{Souza2001-xr} implemented in \texttt{Wannier90} to select the optimal 3-band subspace for constructing MLXWFs.

In Table~\ref{tab:spread_values} we report on the convergence of the Wannierization procedure for both singlet and triplet excitons with increasing $\mathbf{Q}$-mesh. We decompose the total spread, $\Omega$, into its invariant, $\Omega_{\text{I}}$, off-diagonal, $\Omega_{\text{OD}}$, and diagonal, $\Omega_{\text{D}}$, contributions (see Ref.~\cite{Marzari1997-kz} for details on the decomposition). As discussed previously, the underlying $\mathbf{k}$-mesh is chosen twice as dense as the $\mathbf{Q}$-mesh. Despite this additional variability, we still observe clear convergence trends. The convergence of the total spread, $\Omega$, is relatively slow, stemming primarily from the gauge invariant part of the spread, $\Omega_{\text{I}}$, which is fixed after the disentanglement procedure is complete. By contrast the gauge dependent part of the spread, $\Omega_{\text{D}}+\Omega_{\text{OD}}$, the part which is actually minimized after the subspace is fixed, converges rapidly with increasing $\mathbf{Q}$-mesh, changing by less than 0.002 \AA$^2$ between the final two steps. This behavior is analogous, and the level of convergence in $\Omega$ similar, to what was reported at one-electron level in the original paper by Marzari and Vanderbilt~\cite{Marzari1997-kz} in simple semiconducting systems. We further find that $\Omega_{\text{D}}$ is strictly zero, an indication that all Wannier functions possess an inversion center.

\begin{table}[!ht]
    \centering
    \begin{tabular}{c c c c c c }
    \hline
    \hline
        $\mathbf{Q}$-mesh & $\Omega$ & $\Omega_{\text{I}}$ & $\Omega_{\text{OD}}$ & $\Omega_{\text{D}}$ & WF spread  \\
         \hline
         \multicolumn{6}{c}{\textbf{Singlets}} \\
         $3 \times 3 \times 3$ & 4.799 & 4.738 & 0.061 & 0.0 & 1.600 \\
         $4 \times 4 \times 4$ & 5.881 & 5.823 & 0.058 & 0.0 & 1.960 \\
         $5 \times 5 \times 5$ & 6.503 & 6.470 & 0.060 & 0.0 & 2.176 \\
         \multicolumn{6}{c}{\textbf{Triplets}} \\
         $3 \times 3 \times 3$ & 3.980 & 3.957 & 0.023 & 0.0 & 1.327 \\
         $4 \times 4 \times 4$ & 4.494 & 4.476 & 0.018 & 0.0 & 1.498 \\
         $5 \times 5 \times 5$ & 4.797 & 4.778 & 0.017 & 0.0 & 1.599 \\
         \hline 
         \hline
    \end{tabular}
    \caption{Minimized spread $\Omega$ for the triplet excitons in LiF and its decomposition into invariant, $\Omega_{\text{I}}$, off-diagonal, $\Omega_{\text{OD}}$, and diagonal $\Omega_{\text{D}}$ parts. The spread of the individual Wannier functions are reported in the final column. All values are reported in \AA$^2$.}
    \label{tab:spread_values}
\end{table}

In the final column of Table~\ref{tab:spread_values}, we report the spread of the individual exciton Wannier functions, i.e. $\braket{M\bar{\mathbf{0}}|R^2|M\bar{\mathbf{0}}}- \braket{M\bar{\mathbf{0}}|\mathbf{R}|M\bar{\mathbf{0}}}^2$. We find all three Wannier functions have the same spread, stemming from the cubic symmetry of LiF, and in Table~\ref{tab:spread_values} we report a single value in this column. For a given $\mathbf{Q}$-mesh, we find the singlet MLXWFs have larger spreads relative to the triplets. The result is expected as singlets experience an additional exchange force which further delocalizes these states relative to their triplet counterparts.

\subsection{Visualizing the MLXWFs}
\begin{figure}[!ht]
    \centering
    \includegraphics[scale=0.50]{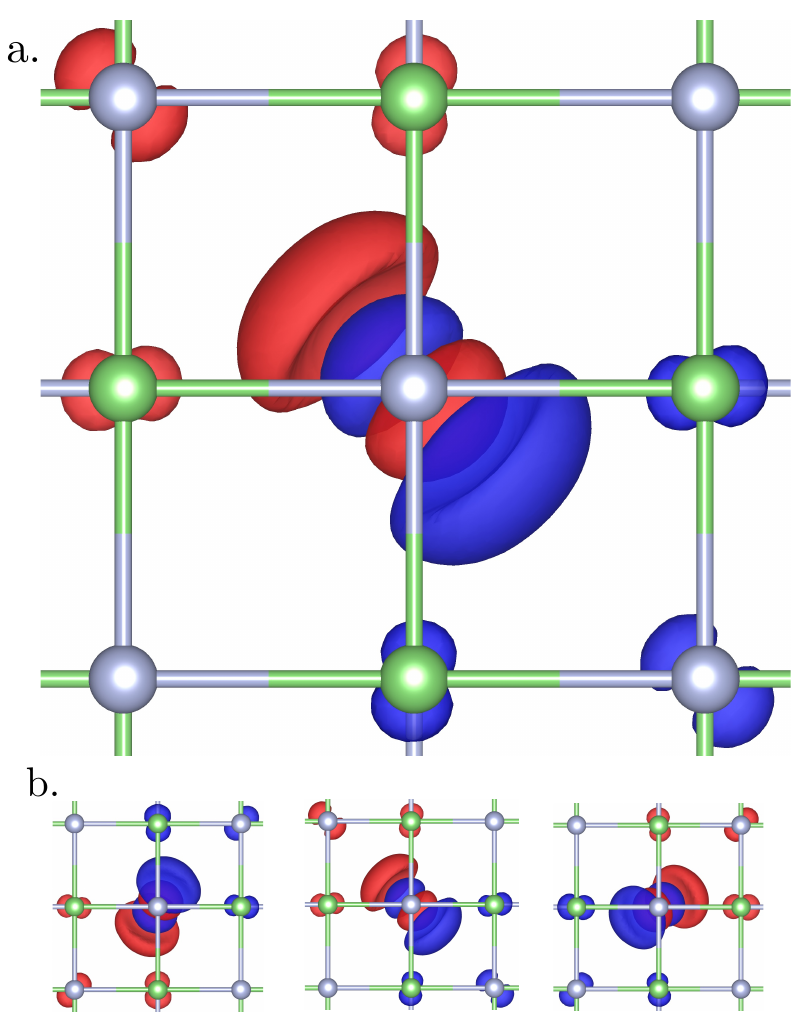}
    \caption{Triplet MLXWF for LiF, plotted as a function of the average coordinate $\mathbf{R}$ with the relative coordinate $\mathbf{r}=\mathbf{0}$ -- i.e. $W_{M\bar{\mathbf{0}}}(\mathbf{R},\mathbf{0})$ vs. $\mathbf{R}$. Green and grey spheres denote Li and F atoms respectively. The Wannier function is entirely real so that blue and red lobes denote regions of positive and negative probability amplitude, respectively. In panel a. we show an enlarged version of the MLXWF for $M=2$. In panel b. we show the three MLXWFs ($M$=1-3).}
    \label{fig:mlxwf_cube}
\end{figure}

\EDIT{To examine the nature of the MLXWF, in Fig.~\ref{fig:mlxwf_cube}, we plot isosurfaces of the MLXWF in $\mathbf{R}$ with $\mathbf{r}$ held fixed at $\mathbf{0}$. Explicitly we plot}
\begin{equation}
    W_{M\bar{\mathbf{0}}}(\mathbf{R},\mathbf{r}=\mathbf{0}) = \frac{1}{\sqrt{N_{\mathbf{Q}}}} \sum_{S\mathbf{Q}}U_{SM}(\mathbf{Q}) \Psi_{S\mathbf{Q}}(\mathbf{R},\mathbf{r}=\mathbf{0})
\end{equation} 
vs. $\mathbf{R}$. \EDIT{ While it is possible more generally to restrict $\mathbf{r}=\mathbf{c}$, where $\mathbf{c}$ is some constant vector, our choice $\mathbf{r}=\mathbf{0}$ is motivated by the fact that it is precisely this component which appears in the exchange integral and MLXWF dipole moments defined in Eqs.~\ref{eq:KX_wann} and~\ref{eq:P_wann}, respectively.} We observe that the MLXWF is well localized in $\mathbf{R}$ in the cell $\bar{\mathbf{R}}=\bar{\mathbf{0}}$. This should be contrasted with $\Psi_{S\mathbf{Q}}(\mathbf{R},\mathbf{r}=\mathbf{0})$\EDIT{, which is delocalized in $\mathbf{R}$ across the entire crystal.} In panel a. we show an enlarged version of the MLXWF. We find that the MLXWF are centered on the fluorine site. The MLXWF has non-trivial nodal structure and in future work it will be interesting to rationalize its shape in terms of linear combinations of electron-hole product states. In panel b. we plot the isosurfaces for the other two MLXWFs. 

In analogy to the one-electron case, our numerical results indicate that the Wannierized excitons can be made entirely real through multiplication by a complex phase. For triplets we find $\text{Im} W / \text{Re} W \approx 0.0001$ while for singlets $\text{Im} W / \text{Re} W \approx 0.01$. Further, we see the MLXWFs transform as odd functions under inversion symmetry and have a non-zero dipole moment. For singlet excitons, this dipole moment gives rise to long-range dipolar interactions between MLXWFs (see Eq.~\ref{eq:KDip_wann}) and drives the splitting of the longitudinal and transverse exciton branches as previously discussed in Sec.~\ref{sec:4B}.

\subsection{MLXWF Dipole Moments and LT Splitting} 
The MLXWF dipole moments can in principle be computed through real space integration of the MLXWF depicted in Fig.~\ref{fig:mlxwf_cube}, following Eq.~\ref{eq:P_wann}. However, here it is more convenient to compute them via Eq.~\ref{eq:PM_practical} using $A^{S\mathbf{0}}_{cv\mathbf{k}}$ and $U_{SM}(\mathbf{0})$, quantities we already have in hand at the end of our BSE and Wannierization calculation, respectively. In Table~\ref{tab:dielectric} we report the magnitude of the MLXWF dipoles.

\begin{table}[!ht]
    \centering
    \begin{tabular}{c c c c c}
    \hline
    \hline
        $\mathbf{k}$-mesh & $N$ & $|\mathbf{P}_M|$ & $\varepsilon_{\text{BG}}(E_{S_{l}})$ & $\EDIT{\varepsilon_{\text{BG}}(E_{S_{t}})}$ \\
         \hline
         $3 \times 3 \times 3$ & 645 & \EDIT{0.260} & 2.06 & \EDIT{1.58} \\
         $4 \times 4 \times 4$ & 1533 & \EDIT{0.261} & 2.06 & \EDIT{1.59}  \\
         $5 \times 5 \times 5$ & 2997 & \EDIT{0.261} & 2.04 &  \EDIT{1.58} \\
         \hline 
         \hline
    \end{tabular}
    \caption{Magnitude of MLXWF dipole moments (in \AA), $|\mathbf{P}_M|$ and the background dielectric constant evaluated at the longitudinal \EDIT{and transverse} exciton energies.}
    \label{tab:dielectric}
\end{table}

The MLXWF dipole moments are the main ingredients for constructing the dipolar interaction between MLXWFs. As discussed in Sec.~\ref{sec:5B}, it is important to further screen this interaction to correctly account of coupling to excitons outside the Wannierization subspace. 

In Table~\ref{tab:dielectric} we report the values of \EDIT{background dielectric evaluated at the longitudinal and transverse exciton energies} computed with different $\mathbf{Q}$-meshes. As the $\mathbf{Q}$-mesh increases, so does the $\mathbf{k}$-mesh, and consequently the number of solutions to the BSE, $N_{\text{Xct}}$. In constructing the effective dielectric function, see Eq.~\ref{eq:dielectric}, we must sum over all states outside Wannierization window $N = N_{\text{Xct}} - N_{\mathcal{\EDIT{\mathcal{W}}}}$. Presently $N_{\EDIT{\mathcal{W}}}$ is held fixed at 3. Despite the increase in the number of states,  $\varepsilon(E_{S_l})\approx 2.0$ \EDIT{and $\varepsilon(E_{S_t})\approx 1.6$} across all $\mathbf{Q}$-meshes, pointing to the stability of the procedure.

\subsection{Exciton Band Structure Interpolation}
As a first application, we show how MLXWFs can be used to interpolate exciton energies throughout the Brillouin zone. Our goal here is to illustrate that the MLXWFs functions can be used as a basis for constructing ab-initio tight-binding models which faithfully reproduces the exciton dispersion and contains the same physics as the usual Bloch description. Additional applications are expounded upon in the following section.

\begin{figure}[!ht]
    \centering
    \includegraphics[scale=0.58]{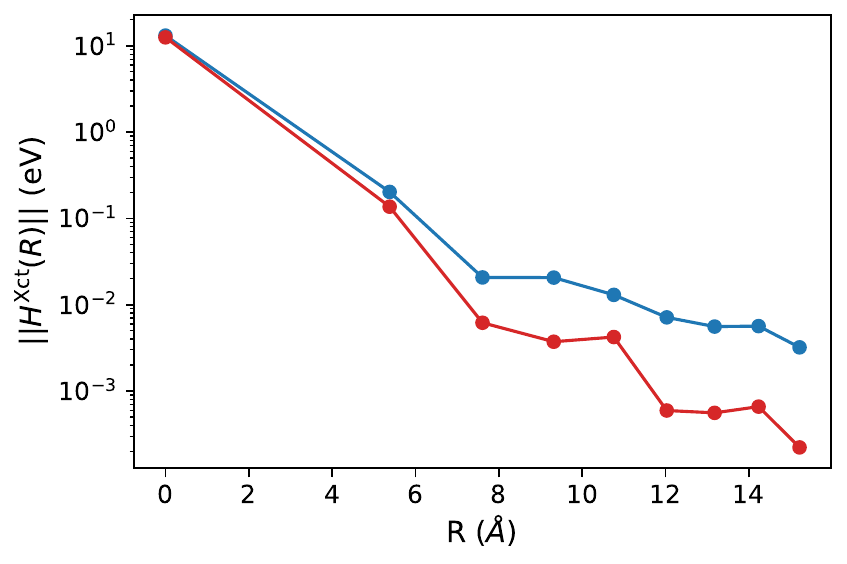}
    \caption{Decay of the singlet (blue curve) and triplet (red curve) BSE Hamiltonian in the Wannier representation $H^{\text{Xct}}_{MN}(\bar{\mathbf{R}})$ as a function of $\bar{\mathbf{R}}$ for LiF. The data points correspond to the largest value of the matrix element at a given $|\mathbf{\bar{R}}|$ -- i.e. $||H(\bar{\mathbf{R}})|| = \max_{MN,|\mathbf{R}|=R}|H^{\text{Xct}}_{MN}(\mathbf{R})|$. } \label{fig:spatial_decay}
\end{figure}

In Fig.~\ref{fig:spatial_decay} we plot the magnitude of the BSE Hamiltonian in the Wannier basis, $H^{\text{Xct}}_{MN}(\bar{\mathbf{R}})$, as a function of $|\bar{\mathbf{R}}|$ in a manner analogous to what is done in  Ref.~\cite{Giustino2007-kc} for $H^{\text{DFT}}$. Physically these matrix elements can be interpreted as hopping matrix elements in the MLXWF basis. As expected, we observe that the triplet matrix elements decay exponentially in $|\bar{\mathbf{R}}|$ while the singlet matrix elements appear to decay more slowly. Theoretically we expect $H^{\text{Xct}}_{MN}(\bar{\mathbf{R}})$ to decay as $1/|\bar{\mathbf{R}}|^3$ for singlets. Our numerical results highlight the importance of removing the long-range dipolar coupling before interpolating.

\begin{figure*}
    \centering
    \includegraphics{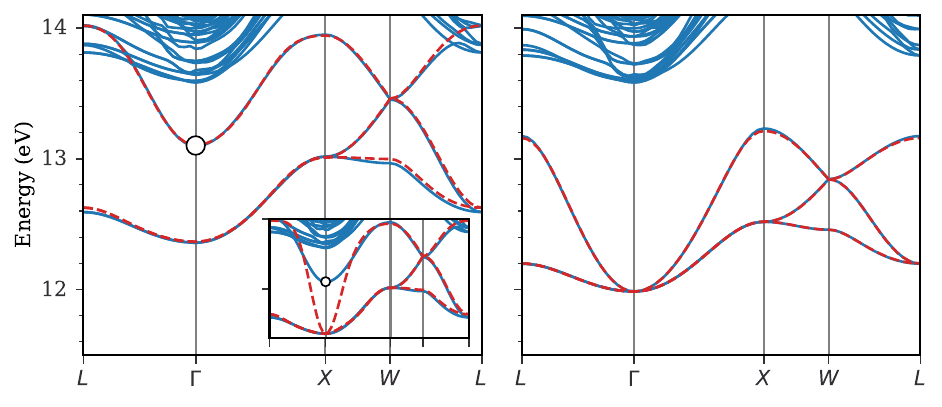}
    \caption{Singlet (left panel) and triplet (right panel) exciton dispersion for LiF. Linear interpolation of exciton eigenergeies explicitly explicitly obtained by diagonalizing the BSE at 72 $\mathbf{Q}$-points along the high-symmetry path (blue curve). Wannier-Fourier interpolated exciton dispersion starting from a $\Gamma$-centered $5\times5\times5$ $\mathbf{Q}$-mesh (dotted red curve). The inset of the left panel shows how Wannier-Fourier interpolation performs without isolating the long-range contribution.}
    \label{fig:exciton_dispersion}
\end{figure*}

In Fig.~\ref{fig:exciton_dispersion} we show the results of our Wannier-Fourier interpolated exciton dispersion for LiF. In blue, we plot a linear interpolation of the singlet and triplet exciton eigenenergies obtained through explicitly diagonalizing the BSE Hamiltonian at 72 $\mathbf{Q}$-points along a high symmetry path. In red we overlay our Wannier-Fourier interpolated exciton dispersion which we computed starting from a $5\times 5 \times 5$ coarse $\mathbf{Q}$-mesh. We find our Wannier-Fourier interpolation scheme (see Sec.~\ref{sec:6}) gives excellent agreement for both the singlet and triplet exciton dispersions. In the inset of the left panel, we show how Wannier-Fourier interpolation preforms without correcting for long-range interactions. In this case, we find that the dispersion agrees relatively well away from $\Gamma$, but completely fails to capture the LT-splitting in the $\mathbf{Q}\rightarrow \mathbf{0}$ limit, highlighting the importance of analytically handling the long-range term.  

\section{Discussion \label{sec:9}}
We expect that MLXWFs could find wide application much as MLWFs have for one-electron states. By design we have formulated our framework to mirror as closely as possible the one-electron case. In doing so we are hopeful that with minimal effort many of the same Wannier-based post-processing tools available for electrons may be readily adapted for the excitonic case. In this work, we have already shown that the internal routines in $\texttt{Wannier90}$ can, with minimal modification, be used to plot MLXWFs and interpolate the exciton band structure. We expect the same to be true for other quantities. For instance, using MLXWFs we can interpolate the exciton-phonon vertex~\cite{Toyozawa1958-dq,Antonius2022-lh,Chen2020-cm} in a manner analogous to how MLWFs are used to interpolate the electron-phonon vertex~\cite{Giustino2007-kc}. Another example is the use of MLXWFs to rotate into a smooth gauge which will facilitate the computation of Berry-phase related properties, like the anomalous hall effect~\cite{Wang2006-hc}, at the excitonic level. 

In adapting these workflows to the excitonic case, it will be important to carefully treat the long-range dipolar coupling between singlet MLXWFs. As previously discussed these long-range couplings are inherently tied to the polarization carried by the exciton, and have no analog at the single electron-level (though there is an analogy with phonons, see Sec.~\ref{sec:4C}). These long-range interactions are responsible for many interesting properties of  excitons. For instance, we have already shown they give rise to splitting of the longitudinal and transverse exciton branches in LiF. In lower symmetry systems these dipolar interactions will lead to non-analytic behavior in the exciton eigenvalues. In the context of exciton-phonon coupling we anticipate a long-range polar coupling between bright excitons and polar phonons. While long-range Fr\"ohlich~\cite{Verdi2015-go} and piezoelectric~\cite{Jhalani2020-fp} electron-phonon coupling have been studied from first-principles, to the best of our knowledge, the polar exciton-lattice coupling mentioned above has yet to be explored. By contrast this long-range coupling has already been linked to the anomalous enhancement of the exciton spin-hall effect relative to the case where electron-hole interactions are neglected~\cite{Onga2017-tk,Kuga2008-jw,Kozin2021-sd}. Our formalism provides a way to compute this enhancement from first-principles.

Beyond Wannier-based post-processing applications we expect the MLXWFs will provide valuable insight into the fundamental nature of the exciton itself. In much the same way that MLWFs provide a bridge between Bloch states and tight-binding orbitals, MLXWFs provide a bridge between the Mott-Wannier~\cite{Mott1938-aw} and Frenkel~\cite{Frenkel1931-qi} starting points for describing an exciton. More specifically, just as the shape of MLWFs can often be rationalized in terms of linear combinations of atomic orbitals, the real space orbital structure of MLXWFs can be rationalized in terms of linear combinations of electron-hole product orbitals. Such analysis could help to better understand the most dominant electron-hole transitions which compose excitons in complex materials. MLXWFs could also serve as a practical tool for quantifying the degree of electron-hole charge-transfer and compliment existing \textit{ab initio} methods~\cite{Sharifzadeh2013-ys}.

MLXWFs also provide an optimal orbital basis for describing the low-energy physics of exciton near the optical band-edge in much the same way that MLWFs provide a efficient basis for describing quasiparticle states near the fundamental band edge, or Fermi surface for metallic systems. Indeed with the quantities computed in this work it is possible to write down an effective tight-binding Hamiltonian for low-lying excitons, namely
\begin{equation} \label{eq:H_eff}
\begin{split}
    H^{\text{Xct}} &= \sum_{\substack{MN \\ \mathbf{R}\mathbf{R}'}} \braket{M\mathbf{0}|H^{\text{SR}}|N\mathbf{D}} c^\dagger_{M\mathbf{R}} c_{N\mathbf{R}'} \\
    &\EDIT{+ \sum_{\substack{MN \\ \mathbf{R}\mathbf{R}'}} 2 \delta_S\braket{M\mathbf{0}|K^{\text{LR,eff}}|N\mathbf{D}}  c^\dagger_{M\mathbf{R}} c_{N\mathbf{R}'}}
\end{split}
\end{equation}
where $\mathbf{D} = \mathbf{R}'-\mathbf{R}$. The first term is readily understood as the usual hopping term while the second term is the dipolar long-range interaction which couples exciton orbitals over long distances. We note that this is not the only partitioning which can be made. For certain applications it may be useful to further isolate higher-order dipole-quadrupole and quadrupole-quadrupole couplings as has recently been done for the electron-phonon vertex~\cite{Jhalani2020-fp}. However as shown in the main text all $\mathbf{Q}$-space non-analyticity is contained in the lowest order dipolar term and for many applications we expect Eq.~\ref{eq:H_eff} may serve as a suitable effective Hamiltonian.

One such application of particular interest is the study of exciton transport. For tightly-bound excitons, the two terms in Eq.~\ref{eq:H_eff} lead to two different transport mechanisms. The first term gives rise to short-range Dexter-like~\cite{Dexter1953-uc} transport while the second term gives rise to long-range Forster-like transport~\cite{Forster1948-mp,Forster1959-jq}. The MLXWF computed in this work can be used as a diabatic basis for studying exciton dynamics~\cite{Jin2017-kb}, shedding light on exciton dynamics.

Finally, in this work, we have generalized the 1-particle Wannierization procedure to handle a specific 2-particle excitation (the exciton). In principle this procedure may readily be adapted to handle other 2-particle excitations, for example the bipolaron, as well as excitations involving more than two particles, for instance, trions and bi-excitons, and beyond.

\section{Conclusions \label{sec:10}} 
In this work, we have developed a procedure for constructing maximally-localized exciton Wannier functions (MLXWFs). We have benchechmarked our work on LiF, and demonstrated that the MLXWFs converge smoothly with \EDIT{increasingly dense} $\mathbf{Q}$-mesh\EDIT{es} and can be used for post-processing applications, like Wannier-Fourier interpolation of the exciton dispersion. We expect this work will serve as a starting point for many other post-processing applications as discussed in Sec.~\ref{sec:9}. Our framework helps to connect the classic Mott-Wannier and \EDIT{tight-binding} pictures of excitons in a first principles context further bridging the gap between condensed matter and quantum chemistry methods for computing neutral excitations. 

\begin{acknowledgments}
This work was supported by the Center for Computational Study of Excited-State Phenomena in Energy
Materials (C2SEPEM)
at the Lawrence Berkeley National Laboratory, funded
by the U.S. Department of Energy, Office of Science,
Basic Energy Sciences, Materials Sciences and Engineering Division, under Contract No. DE-AC02-05CH11231. The development of the screened exchange approach was partially supported by the National Science Foundation (NSF) Condensed Matter and Materials Theory (CMMT) program under Grant DMR-2114081. Computational resources were provided by the National
Energy Research Scientific Computing Center (NERSC).
\end{acknowledgments}

\onecolumngrid
\appendix

\section{Definitions and Conventions \label{app:A}}

\subsection{One-electron states}
We use the following standard definitions and conventions for the single-particle Bloch states 
\begin{equation} 
    \psi_{n\mathbf{k}}(\mathbf{r}) = \frac{1}{\sqrt{N_{\mathbf{k}}}} e^{i\mathbf{k}\cdot\mathbf{r}} u_{n\mathbf{k}}(\mathbf{r})
\end{equation}
and
\begin{equation}
    u_{n\mathbf{k}}(\mathbf{r}) = \frac{1}{\sqrt{V_{\text{uc}}}} \sum_{\mathbf{G}} c_{n\mathbf{k}}(\mathbf{G}) e^{i\mathbf{G}\cdot \mathbf{r}}, \label{eq:u_nk}
\end{equation}
where $u_{n\mathbf{k}}(\mathbf{r})$ is the cell-periodic part of the Bloch state, $\mathbf{G}$ a reciprocal lattice vector, $V_{\text{uc}}$ the volume of the unit cell and $N_{\mathbf{k}}$ the number of $\mathbf{k}$-points in the regular mesh the on which the wavefunctions are computed (equivalently the number of unit cell in the periodic Born–von Karman supercell). Taking the Fourier components $c_{n\mathbf{k}}(\mathbf{G})$ to be orthonormal in $n$ for a fixed $\mathbf{k}$, i.e.
\begin{equation}
    \sum_{\mathbf{G}} c^\star_{n\mathbf{k}}(\mathbf{G}) c_{n'\mathbf{k}}(\mathbf{G}) =\delta_{nn},
\end{equation}
it immediately follows that $u_{n\mathbf{k}}(\mathbf{r})$ and $\psi_{n\mathbf{k}}(\mathbf{r})$ are orthonormal in the unit cell and crystal supercell, respectively, that is
\begin{equation}
    \int_{\text{uc}} d\mathbf{r} u^\star_{n\mathbf{k}}(\mathbf{r}) u_{n'\mathbf{k}}(\mathbf{r}) = \delta_{nn'}, \label{eq:u_ortho}
\end{equation}
and
\begin{equation}
    \int_{V_{\mathbf{k}}} d\mathbf{r} \psi^\star_{n\mathbf{k}}(\mathbf{r}) \psi_{n'\mathbf{k}}(\mathbf{r}) = \delta_{\mathbf{k}\mathbf{k}'} \delta_{nn'}, 
\end{equation}
where the notation $\int_{V_{\mathbf{k}}}$ indicated that the integral is to be performed over $V_{\mathbf{k}} = V_{\text{uc}} N_{\mathbf{k}}$.

\subsection{Exciton States}

\subsubsection{Electron-hole coordinates}

The exciton wavefunction expressed in electron-hole coordinates takes the form
\begin{equation} \label{eq:Psi_rerh}
\begin{split}
    \Psi_{S\mathbf{Q}}(\mathbf{r}_e,\mathbf{r}_h) &= \sum_{cv\mathbf{k}} A^{S
    \mathbf{Q}}_{cv\mathbf{k}} \psi_{c\mathbf{k}}(\mathbf{r}_e) \psi^\star_{v\mathbf{k}-\mathbf{Q}}(\mathbf{r}_h), \\
\end{split}
\end{equation}
where $A^{S\mathbf{Q}}_{cv\mathbf{k}}$ are the exciton expansion coefficients. For a fixed $\mathbf{Q}$, we take the exciton expansion coefficients to be orthonormal in the band index $S$, explicitly
\begin{equation}
\sum_{cv\mathbf{k}} A_{cv\mathbf{k}}^{S'\mathbf{Q}\star} A_{cv\mathbf{k}}^{S\mathbf{Q}} = \delta_{SS'}.
\end{equation}
It follows that the exciton wavefunction as defined in Eq.~\ref{eq:Psi_rerh} is orthonormal in the crystal supercell.

\subsubsection{Weighted-Average and Relative Coordinates}

For the purpose of this work it is convenient re-express the exciton in average $\mathbf{R}$ and relative $\mathbf{r}$ coordinates. In the main text, we defined these coordinates as $\mathbf{R}=(\mathbf{r}_e+\mathbf{r}_h)/2$ and $\mathbf{r}=\mathbf{r}_e-\mathbf{r}_h$. Here we consider the more general transformation for $\mathbf{R}$ given below
\begin{equation} \label{eq:coords}
    \begin{split}
        \mathbf{R} = \alpha \mathbf{r}_e + \beta \mathbf{r}_e,
    \end{split}
\end{equation}
where $\alpha+\beta =1$ and $\alpha,\beta \ge 0$. We now refer to $\mathbf{R}$ as the weighted-average. In this coordinate system, the exciton wavefunction can be written in explicit Bloch periodic form
\begin{equation}
\begin{split}
\Psi_{S\mathbf{Q}}(\mathbf{R},\mathbf{r})  = \frac{1}{\sqrt{N_{\mathbf{Q}}}}e^{i\mathbf{Q}\cdot \mathbf{R}}F_{S\mathbf{Q}}(\mathbf{R},\mathbf{r}),
\end{split}
\end{equation}
where $F_{S\mathbf{Q}}(\mathbf{R},\mathbf{r})$ is the cell-periodic part of the exciton wavefunction and is given by
\begin{equation} \label{eq:FSQ_supp}
\begin{split}
F_{S\mathbf{Q}}(\mathbf{R},\mathbf{r}) =  \frac{1}{\sqrt{N_{\mathbf{k}}}}\sum_{cv\mathbf{k}} A^{S\mathbf{Q}}_{cv\mathbf{k}+\alpha \mathbf{Q}} e^{i\mathbf{k}\cdot\mathbf{r}} u_{c\mathbf{k}+\alpha\mathbf{Q}}(\mathbf{R}+\beta\mathbf{r}) u^\star_{v\mathbf{k}-\beta\mathbf{Q}}(\mathbf{R}-\alpha\mathbf{r}),
\end{split}
\end{equation}
where $N_{\mathbf{Q}}$ and $N_{\mathbf{k}}$ are the number of mesh points used in the $\mathbf{Q}$ and $\mathbf{k}$ meshes, respectively. 

Note by taking $\alpha=\beta=1/2$, $F_{S\mathbf{Q}}(\mathbf{R},\mathbf{r})$ reduces to the specific form reported in Eq.~\ref{eq:FSQ} of the main text. Generally the functional form of $F_{S\mathbf{Q}}(\mathbf{R},\mathbf{r})$ hinges on ones choice of weighted-average coordinate, i.e. the choice of $\alpha$,$\beta$. This dependence is implicit in the $\mathbf{R}$ coordinate. To keep the notation light, we will not explicitly label $F_{S\mathbf{Q}}(\mathbf{R},\mathbf{r})$ with an $\alpha,\beta$ index.

\subsubsection{Normalization}

Below we demonstrate that the exciton wavefunction as defined above is orthonormal in the $\mathbf{k}/\mathbf{Q}$ crystal supercells. By this we mean
\begin{equation}
    \int_{V_{\mathbf{Q}}} d\mathbf{R} \int_{V_{\mathbf{k}}} d\mathbf{r} \Psi^\star_{S\mathbf{Q}}(\mathbf{R},\mathbf{r}) \Psi_{S'\mathbf{Q}'}(\mathbf{R},\mathbf{r}) = \delta_{\mathbf{Q}\mathbf{Q}'} \delta_{SS'},
\end{equation}
where the notation indicates that the integrals over the weighted-average and relative coordinate should be performed over a crystal supercell with volume $V_{\mathbf{Q}} = N_{\mathbf{Q}}V_{\text{uc}}$ and $V_{\mathbf{k}} = N_{\mathbf{k}}V_{\text{uc}}$, respectively. We expect there will be many cases where the $\mathbf{Q}$-mesh needed for Wannierization will be  less dense than the $\mathbf{k}$-mesh on which the BSE must be diagonalized to obtain converged excitonic states. For this reason we delineate between the two meshs at all stages in the following derivations.

To demonstrate orthonormality it is convenient to further partition $F_{S\mathbf{Q}}(\mathbf{R},\mathbf{r})$ as follows
\begin{equation} \label{eq:FSQ_red}
    F_{S\mathbf{Q}}(\mathbf{R},\mathbf{r}) = \sum_{cv\mathbf{k}} A^{S\mathbf{Q}}_{cv\mathbf{k}+\alpha\mathbf{Q}} \chi_{cv\mathbf{k}\mathbf{Q}}(\mathbf{R},\mathbf{r}),
\end{equation}
where
\begin{equation}
    \chi_{cv\mathbf{k}\mathbf{Q}}(\mathbf{R},\mathbf{r}) = \EDIT{\frac{1}{\sqrt{N_{\mathbf{k}}}} e^{i\mathbf{k}\cdot\mathbf{r}} } u_{c\mathbf{k}+\alpha\mathbf{Q}}(\mathbf{R}+\beta\mathbf{r}) u^\star_{v\mathbf{k}-\beta \mathbf{Q}}(\mathbf{R}-\alpha\mathbf{r}).
\end{equation}
\EDIT{and $\chi_{cv\mathbf{k}\mathbf{Q}}(\mathbf{R},\mathbf{r})$ is cell-periodic in $\mathbf{R}$, explicitly $\chi_{cv\mathbf{k}\mathbf{Q}}(\mathbf{R}+\bar{\mathbf{R}},\mathbf{r}) =\chi_{cv\mathbf{k}\mathbf{Q}}(\mathbf{R},\mathbf{r}).$}

\EDIT{
On the way to demonstrating orthonormality of the exciton wavefunctions, it is helpful to prove the following result
\begin{equation}
    \int_{\text{uc}} d\mathbf{R} \int_{V_{\mathbf{k}}} d\mathbf{r} \ \chi^\star_{cv\mathbf{k}\mathbf{Q}}(\mathbf{R},\mathbf{r}) \chi_{c'v'\mathbf{k}'\mathbf{Q}'}(\mathbf{R},\mathbf{r}) = \braket{u_{c\mathbf{k}+\alpha\mathbf{Q}}|u_{c'\mathbf{k}+\alpha\mathbf{Q}'}}_{\text{uc}} \braket{u_{v'\mathbf{k}-\beta\mathbf{Q}'}|u_{v\mathbf{k}-\beta\mathbf{Q}}}_{\text{uc}} \delta_{\mathbf{k}\mathbf{k}'},
\end{equation}
which will be used in Appendix~\ref{app:B}. We make use of the Fourier expansion in Eq.~\ref{eq:u_nk} to express

\begin{equation}
    \chi_{cv\mathbf{k}\mathbf{Q}}(\mathbf{R},\mathbf{r}) = \frac{1}{\sqrt{V_{\mathbf{k}}}} \frac{1}{\sqrt{V_{\text{uc}}}} e^{i\mathbf{k}\cdot\mathbf{r}}\sum_{\mathbf{G}_c,\mathbf{G}_v} c_{c\mathbf{k}+\alpha \mathbf{Q}}(\mathbf{G}_c) c^\star_{v\mathbf{k}-\beta \mathbf{Q}}(\mathbf{G}_v) e^{i\mathbf{G}_c\cdot(\mathbf{R}+\beta \mathbf{r})} e^{-i\mathbf{G}_v\cdot(\mathbf{R}-\alpha \mathbf{r})}.
\end{equation}
Then
\begin{equation} \label{eq:chi_overlap}
\begin{split}
\int_{\text{uc}} d\mathbf{R} \int_{V_{\mathbf{k}}} d\mathbf{r} & \ \chi^\star_{cv\mathbf{k}\mathbf{Q}}(\mathbf{R},\mathbf{r}) \chi_{c'v'\mathbf{k}'\mathbf{Q}'}(\mathbf{R},\mathbf{r}) \\
&= \sum_{\mathbf{G}_c,\mathbf{G}_v,\mathbf{G}_{c'},\mathbf{G}_{v'}} c^\star_{c\mathbf{k}+\alpha\mathbf{Q}}(\mathbf{G}_c) c_{v\mathbf{k}-\beta\mathbf{Q}}(\mathbf{G}_{v}) c_{c'\mathbf{k}'+\alpha\mathbf{Q}'}(\mathbf{G}_{c'}) c^\star_{v'\mathbf{k}'-\beta\mathbf{Q}'}(\mathbf{G}_{v'}) \\
&\qquad \times \frac{1}{V_{\text{uc}}}\int_{\text{uc}} d\mathbf{R} e^{i(-\mathbf{G}_c + \mathbf{G}_v + \mathbf{G}_{c'} - \mathbf{G}_{v'}) \cdot \mathbf{R}} \frac{1}{V_{\mathbf{k}}} \int_{V_{\mathbf{k}}} d\mathbf{r} e^{-i(\mathbf{k}-\mathbf{k}')\cdot \mathbf{r}} e^{i\beta(-\mathbf{G}_c+\mathbf{G}_{c'})\cdot\mathbf{r}} e^{i\alpha(-\mathbf{G}_v+\mathbf{G}_{v'})\cdot\mathbf{r}} \\
&= \sum_{\mathbf{G}_c,\mathbf{G}_v,\mathbf{G}_{c'},\mathbf{G}_{v'}} c^\star_{c\mathbf{k}+\alpha\mathbf{Q}}(\mathbf{G}_c) c_{v\mathbf{k}-\beta\mathbf{Q}}(\mathbf{G}_{v}) c_{c'\mathbf{k}'+\alpha\mathbf{Q}'}(\mathbf{G}_{c'}) c^\star_{v'\mathbf{k}'-\beta\mathbf{Q}'}(\mathbf{G}_{v'}) \delta_{\mathbf{G}_c\mathbf{G}_{c'}} \delta_{\mathbf{G}_v\mathbf{G}_{v'}} \delta_{\mathbf{k} \mathbf{k}'}\\
&= \braket{u_{c\mathbf{k}+\alpha\mathbf{Q}}|u_{c'\mathbf{k}+\alpha\mathbf{Q}'}}_{\text{uc}} \braket{u_{v'\mathbf{k}-\beta\mathbf{Q}'}|u_{v\mathbf{k}-\beta\mathbf{Q}}}_{\text{uc}}\delta_{\mathbf{k}\mathbf{k}}.
\end{split}
\end{equation}
From this result and orthonormality of the cell-periodic states (see Eq.~\ref{eq:u_ortho}), it immediately follows that 
\begin{equation} \label{eq:chi1}
\braket{\chi_{cv\mathbf{k\mathbf{Q}}}|\chi_{c'v'\mathbf{k'\mathbf{Q}}}}  = \delta_{cc'}\delta_{vv'} \delta_{\mathbf{k}\mathbf{k}'}.
\end{equation}
We can show that the exciton wavefunction is normalized to 1 in the crystal supercell
\begin{align*}
    \braket{\Psi_{S\mathbf{Q}}|\Psi_{S'\mathbf{Q}'}} &= \frac{1}{N_{\mathbf{Q}}} \int_{V_{\mathbf{Q}}} d\mathbf{R} e^{-i(\mathbf{Q}-\mathbf{Q}')\cdot \mathbf{R}}   \int_{V_{\mathbf{k}}} d\mathbf{r}  F^\star_{S\mathbf{Q}}(\mathbf{R},\mathbf{r})  F_{S'\mathbf{Q}'}(\mathbf{R},\mathbf{r}) \\
    &= \delta_{\mathbf{Q}\mathbf{Q}'} \int_{\text{uc}} d\mathbf{R} \int_{V_{\mathbf{k}}} d\mathbf{r} F^\star_{S\mathbf{Q}}(\mathbf{R},\mathbf{r})  F_{S'\mathbf{Q}}(\mathbf{R},\mathbf{r}) \\
    &= \delta_{\mathbf{Q}\mathbf{Q}'} \sum_{cv\mathbf{k},c'v'\mathbf{k}} A^{S\mathbf{Q}\star}_{cv\mathbf{k}+\alpha\mathbf{Q}} A^{S'\mathbf{Q}}_{c'v'\mathbf{k}'+\alpha\mathbf{Q}} \bigg[\int_{\text{uc}} d\mathbf{R}  \int_{V_{\mathbf{k}}} \ d\mathbf{r} \chi^\star_{cv\mathbf{k}\mathbf{Q}}(\mathbf{R},\mathbf{r}) \chi_{c'v'\mathbf{k}'\mathbf{Q}}(\mathbf{R},\mathbf{r}) \bigg] \\
    &= \delta_{\mathbf{Q}\mathbf{Q}'}  \sum_{cvc'v'\mathbf{k}} A_{cv\mathbf{k}+\alpha\mathbf{Q}}^{S\mathbf{Q}\star} A^{S'\mathbf{Q}}_{c'v'\mathbf{k}'+\alpha\mathbf{Q}}  \delta_{cc'} \delta_{vv'} \delta_{\mathbf{k}\mathbf{k}'} \\
    &= \delta_{\mathbf{Q}\mathbf{Q}'}\delta_{SS'},
\end{align*}
where we have used the cell-periodicity of $F_{S\mathbf{Q}}(\mathbf{R},\mathbf{r})$ in $\mathbf{R}$ to reduce integral over $V_{\mathbf{Q}}$ to an integral in the unit cell and in passing from line 3 to 4, we have used Eq.~\ref{eq:chi1}.
}

For later use it will also be helpful to note that
\begin{equation} \label{eq:Psi_ortho}
    \begin{split}
        \int_{V_{\mathbf{Q}}} \Psi_{S\mathbf{Q}}(\mathbf{R},\mathbf{0}) d\mathbf{R} = \int_{\text{uc}} F_{S\mathbf{0}}(\mathbf{R},\mathbf{0}) d\mathbf{R} = \frac{1}{\sqrt{N_{\mathbf{k}}}}  \sum_{cv\mathbf{k}} A^{S\mathbf{0}}_{cv\mathbf{k}} \int_{\text{uc}} u_{c\mathbf{k}}(\mathbf{R}) u^\star_{v\mathbf{k}}(\mathbf{R}) d\mathbf{R} = 0.
    \end{split}
\end{equation}

\section{Derivation of Exciton Overlaps \label{app:B}}

Because the exciton wavefunction has a Bloch periodic structure when expressed in $\mathbf{R}+\alpha\mathbf{r}_e+\beta\mathbf{r}_h$, it is possible to Wannierize in $\mathbf{R}=\alpha\mathbf{r}_e+\beta\mathbf{r}_h$ for any $\alpha,\beta \ge 0$ pair which sum to 1. In other words one can minimize the sum of the spreads of the exciton Wannier function, i.e.
\begin{equation} \label{eq:minimizing_functional}
    \Omega^{\text{Xct}}[U;\alpha,\beta] = \sum_M^J [ \braket{M\bar{\mathbf{0}}|R^2|M\bar{\mathbf{0}}} - \braket{M\bar{\mathbf{0}}|\mathbf{R}|M\bar{\mathbf{0}}}^2]
\end{equation}
for any $\mathbf{R} = \alpha\mathbf{r}_e+\beta\mathbf{r}_h$. In the text we choose $\alpha=\beta=1/2$ for its conceptual simplicity and computational tractability. Below we derive the relevant overlaps which must be passed to \texttt{Wannier90} to Wannierize for arbitrary $\mathbf{R}+\alpha\mathbf{r}_e+\beta\mathbf{r}_h$.

Here it is helpful to explicitly label quantities by their $(\alpha,\beta)$ pair. To Wannierize in the coordinate $\mathbf{R}=\alpha\mathbf{r}_e+\beta\mathbf{r}_h$ we must pass to \texttt{Wannier90} overlaps of the form
\begin{equation}
    M^{(\alpha,\beta)}_{SS'}(\mathbf{Q},\mathbf{B}) = \braket{F^{(\alpha,\beta)}_{S\mathbf{Q}}|F^{(\alpha,\beta)}_{S'\mathbf{Q}+\mathbf{B}}}.
\end{equation}
Starting from Eq.~\ref{eq:FSQ_red} we find
\begin{equation} \label{eq:MSS_derivation}
\begin{split}
     M^{(\alpha,\beta)}_{SS'}(\mathbf{Q},\mathbf{B}) &=  \sum_{cv\mathbf{k},c'v'\mathbf{k}'} A^{S\mathbf{Q}\star}_{cv\mathbf{k}\EDIT{+\alpha\mathbf{Q}}} A^{S'\mathbf{Q}+\mathbf{B}}_{c'v'\mathbf{k}'\EDIT{+\alpha\mathbf{Q}}\EDIT{+\alpha\mathbf{B}}} \bigg[ \int_{\text{uc}} d\mathbf{R} \int_{V_\mathbf{k}} d\mathbf{r} \chi^{{(\alpha,\beta)}\star}_{cv\mathbf{k}\mathbf{Q}}(\mathbf{R},\mathbf{r}) \chi^{(\alpha,\beta)}_{c'v'\mathbf{k}'\mathbf{Q}+\mathbf{B}}(\mathbf{R},\mathbf{r}) \bigg] \\
     &= \sum_{cvc'v'\mathbf{k}} A^{S\mathbf{Q}\star}_{cv\mathbf{k}+\alpha\mathbf{Q}} A^{S'\mathbf{Q}+\mathbf{B}}_{c'v'\mathbf{k}\EDIT{'}+\alpha\mathbf{Q}+\alpha\mathbf{B}} \braket{u_{c\mathbf{k}+\alpha\mathbf{Q}}|u_{c'\mathbf{k}+\alpha\mathbf{Q}+\alpha\mathbf{B}}}_{\text{uc}} \braket{u_{v'\mathbf{k}-\beta\mathbf{Q}-\beta\mathbf{B}}|u_{v\mathbf{k}-\beta\mathbf{Q}}}_{\text{uc}} \EDIT{\delta_{\mathbf{k\mathbf{k}'}}} \\
     &= \sum_{cvc'v'\mathbf{k}} A^{S\mathbf{Q}\star}_{cv\mathbf{k}} A^{S'\mathbf{Q}+\mathbf{B}}_{c'v'\mathbf{k}+\alpha\mathbf{B}} \braket{u_{c\mathbf{k}}|u_{c'\mathbf{k}+\alpha\mathbf{B}}}_{\text{uc}} \braket{u_{v'\mathbf{k}-\mathbf{Q}-\beta\mathbf{B}}|u_{v\mathbf{k}-\mathbf{Q}}}_{\text{uc}}.
\end{split}
\end{equation}
In going from line \EDIT{1} to \EDIT{2}, we have made use of Eq.~\ref{eq:chi_overlap} and in the last line we have made a change of variables, $\mathbf{k} \rightarrow \mathbf{k}-\alpha \EDIT{\mathbf{Q}}$. To recover Eq.~\ref{eq:M_SS} in the main text, take $\alpha=\beta=1/2$.

It is important to understand that while different choices of $\alpha,\beta$ will change the functional which is minimized, it may happen that the $U_{SM}(\mathbf{Q})$ which minimize $\Omega^{\text{Xct}}_{\alpha\beta}[U;\alpha,\beta]$ are independent of $\alpha,\beta$. For LiF we have numerically verified that the choices $\alpha=1,\beta=0$, $\alpha=1/2,\beta=1/2$, and $\alpha=0,\beta=1$ all give the same Wannier rotation matrices. Only the spread reported $\Omega[U;\alpha,\beta]$ differs with choice of $\alpha$ and $\beta$. 

For singlet excitons we find it is also necessary to provide \texttt{Wannier90} with a starting guess. Here we
use already Wannierized triplet states as a starting guess for Wannierization of the singlet excitons. To do so, we pass \texttt{Wannier90} the following overlaps
\begin{equation}
    A^{(\mathbf{Q})}_{SM} =  \braket{\Psi_{S\mathbf{Q}}|W_{M\bar{\mathbf{0}}}} = \sum_{cv\mathbf{k}T} A^{S\mathbf{Q}\star}_{cv\mathbf{k}} A^{T\mathbf{Q}}_{cv\mathbf{k}} U_{TM}(\mathbf{Q}).
\end{equation}

\section{Derivation of Direct and Exchange Terms in the Wannier Basis\label{app:C}}

\subsection{Electron-Hole Kernel in the MLXWF Basis} \label{app:C1}

To derive the direct and exchange terms we work in electron-hole coordinates and subsequently transform to average/relative coordinates. Both the exchange and direct electron-hole kernel take the form
\begin{equation}
\begin{split}
    K_{MN}(\bar{\mathbf{R}}) =\int W_{M\bar{\mathbf{0}}}^\star(\mathbf{r}_e,\mathbf{r}_h) K(\mathbf{r}_e,\mathbf{r}_h, \mathbf{r}'_e,\mathbf{r}'_h) W_{N\bar{\mathbf{R}}}(\mathbf{r}'_e,\mathbf{r}'_h) d[\mathbf{r}],
\end{split}
\end{equation}
where $d[\mathbf{r}]=d\mathbf{r}_e d\mathbf{r}_h d\mathbf{r}'_e d\mathbf{r}'_h$ and the integrals are over the supercells associated with the electron and hole $\mathbf{k}$-point meshes. In this real space electron-hole basis, the screened direct term takes the form
\begin{equation}
\begin{split}
K^{\text{D}}(\mathbf{r}_e,\mathbf{r}_h;\mathbf{r}_e,\mathbf{r}'_h) =  \varepsilon^{-1}(\mathbf{r}_e,\mathbf{r}_h) v(\mathbf{r}_e,\mathbf{r}_h) \delta(\mathbf{r}_e,\mathbf{r}'_e) \delta(\mathbf{r}_h,\mathbf{r}'_h),
\end{split}
\end{equation}
while the exchange operator takes the form
\begin{equation}
\begin{split}
K^{\text{X}}(\mathbf{r}_e,\mathbf{r}_h; \mathbf{r}'_e,\mathbf{r}'_h) = v(\mathbf{r}_e,\mathbf{r}_h) \delta(\mathbf{r}_e,\mathbf{r}_h)   \delta(\mathbf{r}'_e,\mathbf{r}'_h).
\end{split}
\end{equation}
Substituting these expressions in we find
\begin{equation}
\begin{split}
K^{\text{D}}_{MN}(\bar{\mathbf{R}}) = \int W^\star_{M\bar{\mathbf{0}}}(\mathbf{r}_e,\mathbf{r}_h) \varepsilon^{-1}(\mathbf{r}_e,\mathbf{r}_h)v(\mathbf{r}_e,\mathbf{r}_h) W_{N\bar{\mathbf{R}}}(\mathbf{r}_e,\mathbf{r}_h) d\mathbf{r}_e d\mathbf{r}_h,  \\
\end{split}
\end{equation}
and
\begin{equation}
\begin{split}
K^{\text{X}}_{MN}(\bar{\mathbf{R}}) = \int W^\star_{M\bar{\mathbf{0}}}(\mathbf{r}_e,\mathbf{r}_e) v(\mathbf{r}_e,\mathbf{r}_h) W_{N\bar{\mathbf{R}}}(\mathbf{r}_h,\mathbf{r}_h) d\mathbf{r}_e d\mathbf{r}_h,  \\
\end{split}
\end{equation}
When re-expressed in the weighted-average  $\mathbf{R}=\alpha\mathbf{r}_e+\beta\mathbf{r}_h$ and relative  
$\mathbf{r}=\mathbf{r}_e-\mathbf{r}_h$ coordinate $W(\mathbf{r}_e,\mathbf{r}_h) \rightarrow W_{M\bar{\mathbf{0}}}(\mathbf{R},\mathbf{r})$ while $W(\mathbf{r}_e,\mathbf{r}_e) \rightarrow W_{M\bar{\mathbf{0}}}(\mathbf{R},\mathbf{0})$. With these substitutions, and adopting the normalization scheme described in Appendix~\ref{app:A}, we find
\begin{equation} \label{eq:KD_wann_supp}
K^{\text{D}}_{MN}(\bar{\mathbf{R}}) = \int_{V_{\mathbf{Q}}} d\mathbf{R} \int_{V_{\mathbf{k}}} d\mathbf{r} \ W_{M\mathbf{0}}^\star(\mathbf{R},\mathbf{r}) \varepsilon^{-1}(\mathbf{R},\mathbf{r})\frac{e^2}{r}W_{N\bar{\mathbf{R}}}(\mathbf{R},\mathbf{r}),
\end{equation}
and
\begin{equation} \label{eq:KX_wann_supp}
K^{\text{X}}_{MN}(\bar{\mathbf{R}}) =  \int_{V_{\mathbf{Q}}} d\mathbf{R} \int_{V_{\mathbf{Q}}} d\mathbf{R}' \ W^\star_{M\mathbf{0}}(\mathbf{R},\bm{0}) \frac{e^2}{|\mathbf{R}-\mathbf{R}'|} W_{N\bar{\mathbf{R}}}(\mathbf{R}',\bm{0}), 
\end{equation}
which are equivalent to Eqs.~\ref{eq:KD_wann} and~\ref{eq:KX_wann} reported in the main text.

\subsection{Leading-Order Expansion of the Exchange Interaction}

Here we provide details on the lowest order expansion of $K^{\text{X}}_{MN}(\bar{\mathbf{R}})$ in $\bar{\mathbf{R}}$.

\subsubsection{Real-space} \label{app:C2A}

We start from the exchange interaction in the MLXWF basis 
\begin{equation} \label{eq:KX_wann_sup}
\begin{split}
K^{\text{X}}_{MN}(\bar{\mathbf{R}}) &=  \int W^\star_{M\mathbf{0}}(\mathbf{R},\bm{0}) \frac{e^2}{|\mathbf{R}-\mathbf{R}'|} W_{N\bar{\mathbf{R}}}(\mathbf{R}',\bm{0}) d\mathbf{R} d\mathbf{R}' \\
 &=  \int W^\star_{M\mathbf{0}}(\mathbf{R},\bm{0}) \frac{e^2}{|\mathbf{R}-\mathbf{R}'-\bar{\mathbf{R}}|} W_{N\bar{\mathbf{0}}}(\mathbf{R}',\bm{0}) d\mathbf{R} d\mathbf{R}',
\end{split}
\end{equation}
where we have made a change of integration $\mathbf{R}' \rightarrow \mathbf{R}' +\bar{\mathbf{R}}$ and used the identity $W_{M\bar{\mathbf{R}}}(\mathbf{R}+\bar{\mathbf{R}},\mathbf{r}) = W_{M\bar{\mathbf{0}}}(\mathbf{R},\mathbf{r})$ to arrive at the second line.

To derive the leading order contribution to this integral we make a \EDIT{multipole} expansion in Coulomb interaction. The lowest order term in this expansion is 
\begin{equation}
    \frac{e^2}{|\mathbf{R}-\mathbf{R}'-\bar{\mathbf{R}}|} = e^2 \bigg[ \frac{\mathbf{R}\cdot\mathbf{R}'}{|\bar{\mathbf{R}}|^3} - 3 \frac{(\mathbf{R}\cdot \bar{\mathbf{R}}) (\mathbf{R}'\cdot \bar{\mathbf{R}})}{|\bar{\mathbf{R}}|^5} \bigg] + \cdots
\end{equation}
Plugging this first order expansion~\ref{eq:KX_wann_sup}, we arrive at the dipolar kernel (see Eq.~\ref{eq:KDip_wann} in the main text)
\begin{equation} \label{eq:KDip_wann_sup}
\begin{split}
&K^{\text{X,Dip}}_{MN}(\bar{\mathbf{R}}) = e^2 \bigg[ \frac{\mathbf{P}^{\star}_M\cdot \mathbf{P}^{}_{N}}{|\bar{\mathbf{R}}|^3} - 3\frac{(\mathbf{P}^{\star}_M \cdot \bar{\mathbf{R}} )(\mathbf{P}^{}_N \cdot  \bar{\mathbf{R}})}{|\bar{\mathbf{R}}|^5} \bigg],
\end{split}
\end{equation}
where
\begin{equation} \label{eq:PM_supp}
    \mathbf{P}_M = \int_{V_{\mathbf{Q}}} W_{M\bar{\mathbf{0}}}(\mathbf{R},\mathbf{0}) \mathbf{R} d\mathbf{R}.
\end{equation}

\subsubsection{Reciprocal space} \label{app:C2B}

In Eq.~\ref{eq:KDip_wann_Q} of the main text we give the Fourier transform of the dipolar interaction.  To derive this result we start with the Fourier expansion of the Coulomb interaction
\begin{equation} \label{eq:coul_four}
\begin{split}
\frac{e^2}{|\mathbf{R}-\mathbf{R}'-\bar{\mathbf{R}}|} = \frac{4\pi e^2}{V_{\mathbf{q}}} \sum_{\mathbf{q}\mathbf{G}} e^{+i(\mathbf{q}+\mathbf{G})\cdot\mathbf{R}} \frac{1}{|\mathbf{q}+\mathbf{G}|^2} e^{-i(\mathbf{q}+\mathbf{G})\cdot\mathbf{R}'} e^{-i\mathbf{q}\cdot\bar{\mathbf{R}}}, \\
\end{split}
\end{equation}
where $V_{\mathbf{q}}=N_{\mathbf{q}}V_{\text{uc}}$ and we have used $e^{i\mathbf{G}\cdot \bar{\mathbf{R}}} = 1$. 

Multiplying both sides $e^{i\mathbf{Q}\cdot\bar{\mathbf{R}}}$ and summing over $\bar{\mathbf{R}}$, we find
\begin{equation}
\begin{split}
\sum_{\bar{\mathbf{R}}}\frac{e^2}{|\mathbf{R}-\mathbf{R}'-\bar{\mathbf{R}}|} e^{i\mathbf{Q}\cdot\bar{\mathbf{R}}} = \frac{4\pi e^2}{V_{\text{uc}}} \sum_{\mathbf{G}} e^{+i(\mathbf{q}+\mathbf{G})\cdot\mathbf{R}} \frac{1}{|\mathbf{Q}+\mathbf{G}|^2} e^{-i(\mathbf{Q}+\mathbf{G})\cdot\mathbf{R}'}, \\
\end{split}
\end{equation}
where we have used $\sum_{\bar{\mathbf{R}}} e^{i(\mathbf{Q}-\mathbf{q})\cdot\bar{\mathbf{R}}} = N_{\mathbf{Q}} \delta_{\mathbf{Q}\mathbf{q}}$.

Plugging Eq.~\ref{eq:coul_four} into Eq.~\ref{eq:KX_wann_sup} we obtain
\begin{align*}
\sum_{\bar{\mathbf{R}}} \bigg(\int & W^\star_{M\bar{\mathbf{0}}}(\mathbf{R},\mathbf{0})  \frac{e^2}{|\mathbf{R}-\mathbf{R}'-\bar{\mathbf{R}}|}  W_{N\bar{\mathbf{0}}}(\mathbf{R}',\mathbf{0}) d\mathbf{R} d\mathbf{R'} \bigg) e^{i\mathbf{Q}\cdot\bar{\mathbf{R}}} \\
&= \frac{4\pi e^2}{V_{\text{uc}}} \sum_{\mathbf{G}} \bigg( \int d\mathbf{R} W^\star_{M\bar{\mathbf{0}}}(\mathbf{R},\mathbf{0})   e^{i(\mathbf{Q}+\mathbf{G})\cdot\mathbf{R}} \bigg) \frac{1}{|\mathbf{Q}+\mathbf{G}|^2} \bigg( \int d\mathbf{R}' e^{-i(\mathbf{Q}+\mathbf{G})\cdot\mathbf{R}'} W_{N\bar{\mathbf{0}}}(\mathbf{R}',\mathbf{0})  \bigg).
\end{align*}

To arrive at the dipolar contribution we expand both sides to lowest order in $\mathbf{R}$ and $\mathbf{R}'$. We find that
\begin{align*}
e^2 \sum_{\bar{\mathbf{R}}} \bigg[ \frac{\mathbf{P}^\star_M \cdot \mathbf{P}_N}{|\bar{\mathbf{R}}|^3} 
 -\frac{(\mathbf{P}^\star_M\cdot\bar{\mathbf{R}})(\mathbf{P}_N\cdot\bar{\mathbf{R}})}{|\bar{\mathbf{R}}|^5}  \bigg] e^{i\mathbf{Q}\cdot\bar{\mathbf{R}}} 
= 
\frac{4\pi e^2}{V_{\text{uc}}} \sum_{\mathbf{G}} \frac{[ \mathbf{P}^\star_M \cdot (\mathbf{Q}+\mathbf{G}) ][\mathbf{P}_N \cdot (\mathbf{Q}+\mathbf{G})]}{|\mathbf{Q}+\mathbf{G}|^2},
\end{align*}
where MLXWF dipolemoments, $\mathbf{P}_M$, are defined in Eq.~\ref{eq:PM_supp}.

\subsection{Exciton Transition Dipole Matrix Elements and Their Relations} \label{app:C3}

The MLXWF transition dipole matrix elements naturally appear when making the multipole expansion of the exchange interaction. Here we discuss properties of these dipole matrix elements and their relation to the usual dipole matrix elements used to study optical properties.

In our multipole expansion we found that only the MLXWF dipoles at $\bar{\mathbf{R}}=\bar{\mathbf{0}}$ were required. Nevertheless in the discussion below it is helpful to consider the MLXWF dipole matrix element at non-zero $\bar{\mathbf{R}}$, defined as
\begin{equation}
    \mathbf{P}_{M\bar{\mathbf{R}}} = \int_{V_{\mathbf{Q}}} \mathbf{R}W_{M\bar{\mathbf{R}}}(\mathbf{R},\mathbf{r}=\mathbf{0})  d\mathbf{R}.
\end{equation}
We also introduce the Bloch exciton dipole matrix element at non-zero $\mathbf{Q}$, defined as
\begin{equation}
    \mathbf{P}_{S\mathbf{Q}} = \int_{V_{\mathbf{Q}}} \mathbf{R} \Psi_{S\mathbf{Q}}(\mathbf{R},\mathbf{r}=\mathbf{0})  d\mathbf{R}.
\end{equation}
Below we will prove
\begin{align}
    \mathbf{P}_{S\mathbf{Q}} &= \mathbf{P}_{S\mathbf{0}}\delta_{\mathbf{Q}\mathbf{0}}, \label{eq:PSQ_delta} \\
    \mathbf{P}_{M\bar{\mathbf{R}}} &= \mathbf{P}_{M\bar{\mathbf{0}}}. \label{eq:PMR_PM0} 
\end{align}
These results can be derived by considering how the Bloch and Wannier exciton dipole moments transform under translation in the weighted-average coordinate.

For the Bloch exciton dipole moment we have 
\begin{equation} \label{eq:PSQ0}
\begin{split}
\mathbf{P}_{S\mathbf{Q}} &= \int_{V_{\mathbf{Q}}} \mathbf{R} \Psi_{S\mathbf{Q}}(\mathbf{R},\mathbf{0})  \ d\mathbf{R} \\
&= \int_{V_{\mathbf{Q}}} (\mathbf{R}+\bar{\mathbf{R}}) \Psi_{S\mathbf{Q}}(\mathbf{R}+\bar{\mathbf{R}},\mathbf{0}) \ d\mathbf{R} \\
&=\int_{V_{\mathbf{Q}}} (\mathbf{R}+\bar{\mathbf{R}}) \Psi_{S\mathbf{Q}}(\mathbf{R},\mathbf{0}) e^{i\mathbf{Q}\cdot\bar{\mathbf{R}}} \ d\mathbf{R} \\
&= e^{i\mathbf{Q}\cdot\bar{\mathbf{R}}} \int_{V_{\mathbf{Q}}} \mathbf{R}\Psi_{S\mathbf{Q}}(\mathbf{R},\mathbf{0})  d\mathbf{R} + \bar{\mathbf{R}} e^{i\mathbf{Q}\cdot\bar{\mathbf{R}}} \int_{V_{\mathbf{Q}}} \Psi_{S\mathbf{Q}}(\mathbf{R},\mathbf{0}) \ d\mathbf{R} \\
&= e^{i\mathbf{Q}\cdot\bar{\mathbf{R}}} \int_{V_{\mathbf{Q}}} \mathbf{R}\Psi_{S\mathbf{Q}}(\mathbf{R},\mathbf{0})  d\mathbf{R} +\mathbf{0} \\
& =e^{i\mathbf{Q}\cdot\bar{\mathbf{R}}} \mathbf{P}_{S\mathbf{Q}},
\end{split}
\end{equation}
where we have shifted the integration variable by a lattice constant, applied Bloch's theorem, and used Eq.~\ref{eq:Psi_ortho} to arrive at the final line. To complete the argument, note that the only way for $\mathbf{P}_{S\mathbf{Q}}=e^{i\mathbf{Q}\cdot\bar{\mathbf{R}}} \mathbf{P}_{S\mathbf{Q}}$  to hold for arbitrary $\bar{\mathbf{R}}$ is for $\mathbf{P}_{S\mathbf{Q}} = \mathbf{P}_{S\mathbf{0}}\delta_{\mathbf{Q}\mathbf{0}}$ which proves Eq.~\ref{eq:PSQ_delta}. 

Along similar lines, we have

\begin{equation} \label{eq:P_trans}
\begin{split}
    \mathbf{P}_{M\bar{\mathbf{R}}} &= \int_{V_{\mathbf{Q}}} \mathbf{R} W_{M\bar{\mathbf{R}}}(\mathbf{R},\mathbf{0}) d\mathbf{R} \\
    &= \int_{V_{\mathbf{Q}}} (\mathbf{R}+\bar{\mathbf{R}}) W_{M\bar{\mathbf{R}}}(\mathbf{R}+\bar{\mathbf{R}},\mathbf{0}) d\mathbf{R} \\
    &= \int_{V_{\mathbf{Q}}} \mathbf{R} W_{M\bar{\mathbf{0}}}(\mathbf{R},\mathbf{0}) d \mathbf{R}  + \bar{\mathbf{R}}\int_{V_{\mathbf{Q}}} W_{M\bar{\mathbf{0}}}(\mathbf{R},\mathbf{0}) d\mathbf{R} \\
    &= \int_{V_{\mathbf{Q}}} \mathbf{R} W_{M\bar{\mathbf{0}}}(\mathbf{R},\mathbf{0}) d \mathbf{R} + \mathbf{0}\\
    &=\mathbf{P}_{M\bar{\mathbf{0}}},
\end{split}
\end{equation}
where we have used the relation $W_{M\bar{\mathbf{R}}}(\mathbf{R}+\bar{\mathbf{R}},\mathbf{r}) = W_{M\bar{\mathbf{0}}}(\mathbf{R},\mathbf{r})$ and $\int W_{M\bar{\mathbf{0}}}(\mathbf{R},\mathbf{r}=0) d\mathbf{R} =0$, this last relation follows from Eq.~\ref{eq:Psi_ortho}.

With these results we can show that the Bloch and MLXWF transition dipole matrix elements are related through the Wannier rotation matrices at $\mathbf{Q}=\mathbf{0}$. We start from
\begin{equation} \label{eq:PsiS0}
\begin{split}
    \Psi_{S\mathbf{0}}(\mathbf{R},\mathbf{r}=\mathbf{0}) = \frac{1}{\sqrt{N_{\mathbf{Q}}}} \sum_{M\bar{\mathbf{R}}} U^\dagger_{MS}(\mathbf{0}) W_{M\bar{\mathbf{R}}}(\mathbf{R},\mathbf{r}=\mathbf{0}), 
\end{split}
\end{equation}
which follows from Eq.~\ref{eq:xct_mlwf_inv}, and integrate both sides against $\mathbf{R}$ to find
\begin{equation} \label{eq:PM2PS}
\begin{split}
    \mathbf{P}_{S\mathbf{0}} &= \frac{1}{\sqrt{N_{\mathbf{Q}}}} \sum_{M\bar{\mathbf{R}}} U^\dagger_{MS}(\mathbf{0}) P_{M\bar{\mathbf{R}}} = \sqrt{N_{\mathbf{Q}}} \sum_{M} U^\dagger_{MS}(\mathbf{0}) \mathbf{P}_{M\bar{\mathbf{0}}},
\end{split}
\end{equation}
where we have used $\mathbf{P}_{M\bar{\mathbf{R}}}=\mathbf{P}_{M\bar{\mathbf{0}}}$ (see Eq.~\ref{eq:PMR_PM0}) and the trivial relation $\sum_{\bar{\mathbf{R}}}1=N_{\mathbf{Q}}$. Eq.~\ref{eq:PM2PS} can be directly inverted to express $\mathbf{P}_{M\bar{\mathbf{0}}}$ in terms of $\mathbf{P}_{S\mathbf{0}}$. Alternatively, we can derive the inverse relation starting from
\begin{equation}  \label{eq:WM0}
\begin{split}
    W_{M\bar{\mathbf{0}}}(\mathbf{R},\mathbf{r}=\mathbf{0}) = \frac{1}{\sqrt{N_{\mathbf{Q}}}} \sum_{S\mathbf{Q}} U_{SM}(\mathbf{Q}) \Psi_{S\mathbf{Q}}(\mathbf{R},\mathbf{r}=\mathbf{0}), 
\end{split}
\end{equation}
which follow from Eq.~\ref{eq:xct_mlwf} in the main text. Again we integrate both sides with respect to $\mathbf{R}$ to find
\begin{equation} \label{eq:PS2PM}
\begin{split}
    \mathbf{P}_{M\bar{\mathbf{0}}} = \frac{1}{\sqrt{N_{\mathbf{Q}}}} \sum_{S\mathbf{Q}} U_{SM}(\mathbf{Q}) \mathbf{P}_{S\mathbf{Q}} = \frac{1}{\sqrt{N_{\mathbf{Q}}}} \sum_S U_{SM}(\mathbf{0}) \mathbf{P}_{S\mathbf{0}},
\end{split}
\end{equation}
where we have used $\mathbf{P}_{S\mathbf{Q}}=\mathbf{P}_{S\mathbf{0}} \delta_{\mathbf{Q}\mathbf{0}}$ (see Eq.~\ref{eq:PSQ_delta}). Eq.~\ref{eq:PM2PS} and Eq.~\ref{eq:PS2PM} coincide with Eq.~\ref{eq:P_bloch2wann} in the main text.

\subsection{Exciton Position Matrix Element}

The exciton transition dipole matrix elements should be distinguished from the expectation value of the position operator. The latter, in the MLXWF basis, is defined as
\begin{equation}
\begin{split}
\braket{\mathbf{R}}_{M\bar{\mathbf{R}}} &=\int_{V_{\mathbf{Q}}} d\mathbf{R} \int_{V_{\mathbf{k}}} d\mathbf{r} W^\star_{M\bar{\mathbf{R}}}(\mathbf{R},\mathbf{r}) \mathbf{R} W_{M\bar{\mathbf{R}}}(\mathbf{R},\mathbf{r}) .
\end{split}
\end{equation}
Importantly the position operator matrix elements of MLXWFs centered on different cells are related by the addition of a lattice vector $\bar{\mathbf{R}}$. Explicitly,
\begin{equation} \label{eq:R_trans}
    \braket{\mathbf{R}}_{M\bar{\mathbf{R}}} = \braket{\mathbf{R}}_{M} + \bar{\mathbf{R}}.
\end{equation}
This relation is analogous to what is seen at the one-electron level and plays an important role in many post-processing applications where a consistent choice of $\bar{\mathbf{R}}$ is crucial to obtain correct results, e.g., the Berry phase theory of polarization. Eq.~\ref{eq:R_trans} should be contrasted with Eq.~\ref{eq:P_trans}, notably the dipole matrix elements are invariant under translation. 

\section{Derivation of the Effective Non-Analytic kernel} \label{app:D}

\EDIT{

\subsubsection{Downfolded non-analytic kernel}

In this section we derive the downfolded non-analytic kernel (see Eq.~\ref{eq:KNA_eff}). Our starting point is the infinite series in Eq.~\ref{eq:RSPT}

\begin{equation} \label{eq:RSPT_app}
K^{\text{NA},\text{eff}}_{SS'}(\mathbf{Q}) = K^{\text{NA}}_{SS'}(\mathbf{Q}) + 2\sum_{R\notin \mathcal{W}}\frac{K^{\text{NA}}_{SR}(\mathbf{Q})K^{\text{NA}}_{RS'}(\mathbf{Q})}{\omega-E_{R\mathbf{0}}} +2^2\sum_{R,R'\notin \mathcal{W}}\frac{K^{\text{NA}}_{SR}(\mathbf{Q})K^{\text{NA}}_{RR'}(\mathbf{Q})K^{\text{NA}}_{R'S'}(\mathbf{Q})}{(\omega-E_{R\mathbf{0}})(\omega-E_{R'\mathbf{0}})} + \cdots ,
\end{equation}
where $K_{SS'}^{\text{NA}}(\mathbf{Q})$, given in Eq.~\ref{eq:KNA_bloch}, is reported here for clarity
\begin{equation} \label{eq:KNA_app}
K_{SS'}^{\text{NA}}(\mathbf{Q}) = \frac{4 \pi e^2}{ V_{\text{uc}}N_{\mathbf{Q}}} \frac{[\mathbf{P}^\star_{S} \cdot \mathbf{Q}] [\mathbf{P}_{S'} \cdot \mathbf{Q}]}{|\mathbf{Q}|^2}.
\end{equation}
The separable form of $K_{SS'}^{\text{NA}}(\mathbf{Q})$, allows Eq.~\ref{eq:RSPT_app} to be summed to infinite order. We find 
\begin{equation} \label{eq:geo_series}
K^{\text{NA},\text{eff}}_{SS'}(\mathbf{Q}) = K^{\text{NA}}_{SS'}(\mathbf{Q})[1+2v(\mathbf{Q}) \chi^{\text{BG}}(\mathbf{Q}) + [2v(\mathbf{Q}) \chi^{\text{BG}}(\mathbf{Q})]^2+\cdots],
\end{equation}
where the common ratio
\begin{equation}
    2v(\mathbf{Q}) \chi_{\text{BG}}(\mathbf{Q}) = 2\frac{4\pi e^2}{V_{\text{uc}}N_{\mathbf{Q}}|\mathbf{Q}|^2} \sum_{R\notin \mathcal{W}} \frac{[\mathbf{P}_{R} \cdot \mathbf{Q}][\mathbf{P}^\star_{R} \cdot\mathbf{Q}]}{\omega-E_{R\mathbf{0}}},
\end{equation}
can physically be interpreted as the product of the Coulomb interaction, $v(\mathbf{Q})$, and background susceptibility, $\chi_{\text{BG}}(\mathbf{Q})$, in the $\mathbf{Q} \rightarrow \mathbf{0}$ limit. Eq.~\ref{eq:geo_series} sums to
\begin{equation}
K^{\text{NA},\text{eff}}_{SS'}(\mathbf{Q}) =  \varepsilon^{-1}_{\text{BG}}(\mathbf{Q},\omega) K^{\text{NA}}_{SS'}(\mathbf{Q}),
\end{equation}
where 
\begin{equation}
\begin{split}
    \varepsilon_{\text{BG}}(\mathbf{Q},\omega) = 1 - 2 v(\mathbf{Q})\chi_{\text{BG}}(\mathbf{Q},\omega).
\end{split}
\end{equation}
We see that states in the passive space dynamically screen the non-analytic coupling between active space states.

\EDIT{

\subsubsection{Static non-analytic kernel}

In this section we show how to cast the frequency-dependent effective Hamiltonian (see Eq.~\ref{eq:H_downfolded_MN}) as a static non-hermitian Hamiltonian (see Eqs.~\ref{eq:H_downfolded_static}-\ref{eq:epsilon_MN}) in the MLXWF basis. Our starting point is Eq.~\ref{eq:H_downfolded_MN}, reproduced here for clarity
\begin{equation} \label{eq:H_downfolded_MN_app}
\lim_{\mathbf{Q}\rightarrow \mathbf{0}} H^{\text{Xct,eff}}_{MN}(\mathbf{Q},\omega) = H^{\text{SR}}_{MN}(\mathbf{0}) + 2\delta_S K_{MN}^{\text{NA,eff}}(\mathbf{Q},\omega),
\end{equation}
where
\begin{equation} \label{eq:KNA_eff_MN_app}
\begin{split}
    K^{\text{NA,eff}}_{MN}(\mathbf{Q},\omega)
    = \varepsilon_{\text{BG}}^{-1}(\mathbf{Q},\omega) K_{MN}^{\text{NA}}(\mathbf{Q})
\end{split}
\end{equation}
with $K_{MN}^{\text{NA}}(\mathbf{Q})$ defined in Eq.~\ref{eq:KNA_wann_Q} in the main text. Let $E_{S\mathbf{Q}}$ and $C_{MS}(\mathbf{Q})$ denote the eigensolutions of Eq.~\ref{eq:H_downfolded_MN_app} so that
\begin{equation} \label{eq:intermediate_1}
    \sum_{N} \lim_{\mathbf{Q}\rightarrow \mathbf{0}} H^{\text{NA,eff}}_{MN}(\mathbf{Q},E_{S\mathbf{Q}}) C_{NS}(\mathbf{Q}) = E_{S\mathbf{Q}} C_{MS}(\mathbf{Q}).
\end{equation}
Define 
\begin{equation} \label{eq:intermediate_3}
\begin{split}
     \lim_{\mathbf{Q}\rightarrow \mathbf{0}} H^{\text{Xct,eff}}_{MN}(\mathbf{Q}) =  \sum_{N'} \sum_{S'} \lim_{\mathbf{Q}\rightarrow \mathbf{0}} H^{\text{NA,eff}}_{MN'}(\mathbf{Q},E_{S'\mathbf{Q}}) C_{N'S'}(\mathbf{Q}) C^{-1}_{S'N}(\mathbf{Q}) 
\end{split}
\end{equation}
Note that by construction $\lim_{\mathbf{Q}\rightarrow \mathbf{0}} H^{\text{Xct,eff}}_{MN}(\mathbf{Q})$ has the same eigenvalues and eigenvectors as Eq.~\ref{eq:intermediate_1}.

Plugging Eqs.~\ref{eq:H_downfolded_MN_app}  and~\ref{eq:KNA_eff_MN_app} into Eq.~\ref{eq:intermediate_3} we find
\begin{equation} \label{eq:H_downfolded_static_app}
    \lim_{\mathbf{Q}\rightarrow \mathbf{0}} H^{\text{Xct,eff}}_{MN}(\mathbf{Q}) = H^{\text{SR}}_{MN}(\mathbf{0}) + 2\delta_S K_{MN}^{\text{NA,eff}}(\mathbf{Q}),
\end{equation}
where
\begin{equation}
    K^{\text{NA},\text{eff}}_{MN}(\mathbf{Q}) = \sum_{N'}  K^{\text{NA}}_{MN'}(\mathbf{\mathbf{Q}}) [\varepsilon^{-1}_{\text{BG}}(\mathbf{Q})]_{N'N},
\end{equation}
and
\begin{equation} \label{eq:epsilon_MN_app}
    [\varepsilon^{-1}_{\text{BG}}(\mathbf{Q})]_{N'N} = \sum_{S} C_{N'S}(\mathbf{Q}) \varepsilon^{-1}_{\text{BG}}(\mathbf{Q},E_{S\mathbf{Q}})  C^{-1}_{SN}(\mathbf{Q}).
\end{equation}
Eqs.~\ref{eq:H_downfolded_static_app}-\ref{eq:epsilon_MN_app} are equivalent to those given in the main text (see Eqs.~\ref{eq:H_downfolded_static}-\ref{eq:epsilon_MN}).

Note, because $C_{SN}(\mathbf{Q})$ are obtained by diagonalizing Eq.~\ref{eq:H_downfolded_MN_app} they are unique only up to a phase. For example if $C_{NS}(\mathbf{Q})$ satisfies Eq.~\ref{eq:H_downfolded_MN_app} so will $e^{i\phi_S(\mathbf{Q})}C_{NS}(\mathbf{Q})$. In Eq.~\ref{eq:epsilon_MN_app} the arbitrary phases associated with  $C_{N'S}(\mathbf{Q})$ and $C^{-1}_{SN}(\mathbf{Q})$ cancel so that $[\varepsilon^{-1}_{\text{BG}}(\mathbf{Q})]_{N'N}$ is well defined. 
}
}

\section{Ewald Summation \label{app:E}}

\EDIT{

The reciprocal space dipolar kernel, $K^{\text{X,Dip}}(\mathbf{Q})$, defined in Eq.~\ref{eq:KDip_wann_Q} as 
\begin{equation} 
\begin{split}
K^{\text{X,Dip}}_{MN}&(\mathbf{Q}) = \frac{4 \pi e^2}{V_{\text{\rm{uc}}}} \sum_{\mathbf{G}} \frac{ [\mathbf{P}^{\star}_M \cdot (\mathbf{Q}+\mathbf{G}) ] [\mathbf{P}^{}_N \cdot (\mathbf{Q}+\mathbf{G}) ]}{|\mathbf{Q}+\mathbf{G}|^2},  \\
\end{split}
\end{equation}
formally diverges. This divergence stems from the self-interaction ($\bar{\mathbf{R}}=\bar{\mathbf{0}}$) term in the real space dipolar kernel, $K^{\text{X,Dip}}(\bar{\mathbf{R}})$ (see Eq.~\ref{eq:KDip_wann}). By contrast the long-range kernel, defined in Eq.~\ref{eq:K_lr_mn} as 
\begin{equation}  \label{eq:Klr_mn_app}
\begin{split}
    K^{\text{LR}}_{MN}(\mathbf{Q}) =\frac{4 \pi e^2}{V_{\text{\rm{uc}}}} \sum_{\mathbf{G}} \frac{ [\mathbf{P}^{\star}_M \cdot (\mathbf{Q}+\mathbf{G}) ] [\mathbf{P}^{}_N \cdot (\mathbf{Q}+\mathbf{G}) ]}{|\mathbf{Q}+\mathbf{G}|^2} -\frac{4 \pi e^2}{V_{\text{\rm{uc}}}} \sum_{\mathbf{G}\ne \mathbf{0}} \frac{ [\mathbf{P}^{\star}_M \cdot \mathbf{G} ] [\mathbf{P}^{}_N \cdot \mathbf{G} ]}{|\mathbf{G}|^2},
\end{split}
\end{equation}
is convergent. Still, the convergence with respect to the $\mathbf{G}$ sum is slow, a consequence of subtracting two divergent quantities from one another. To efficiently evaluate $K_{MN}^{\text{LR}}(\mathbf{Q})$, we make use of Ewald's technique~\cite{Born1956-ty}. We partition $\smash{K^{\text{X,Dip}}_{MN}(\mathbf{Q})}$ into a rapidly converging reciprocal and real-space sum (see for instance Eq. 73 of Ref.~\cite{Gonze1997-mx} for additional details on this partitioning). 

For a sufficiently large range separation parameter, $\Lambda$, the real-space contribution to the Ewald sum can be made negligible. Retaining only the reciprocal space contribution we arrive at the following approximation to $K_{MN}^{\text{LR}}(\mathbf{Q})$
\begin{equation} \label{eq:Klr_ewald}
\begin{split}
    K_{MN}^{\text{LR}}(\mathbf{Q}) =\frac{4 \pi e^2}{V_{\text{\rm{uc}}}} \bigg[ \sum_{\mathbf{G}} \frac{ [\mathbf{P}^{\star}_M \cdot (\mathbf{Q}+\mathbf{G}) ] [\mathbf{P}^{}_N \cdot (\mathbf{Q}+\mathbf{G}) ]}{|\mathbf{Q}+\mathbf{G}|^2} e^{-(\mathbf{Q}+\mathbf{G})^2/(4\Lambda^2)} - \sum_{\mathbf{G}\ne \mathbf{0}} \frac{ [\mathbf{P}^{\star}_M \cdot \mathbf{G} ] [\mathbf{P}^{}_N \cdot \mathbf{G} ]}{|\mathbf{G}|^2} e^{-(\mathbf{G})^2/(4\Lambda^2)}\bigg].
\end{split}
\end{equation}
Note that for $\Lambda \rightarrow \infty$ Eq.~\ref{eq:Klr_ewald} reduces to~\ref{eq:Klr_mn_app} and is exact. For sufficiently large $\Lambda$, Eq.~\ref{eq:Klr_ewald} gives a very good approximation to Eq.~\ref{eq:Klr_mn_app} and converges much faster, so that only a few $\mathbf{G}$-shells need to be included. For LiF we take $\Lambda$ to be one inverse Bohr.

}

\twocolumngrid

\bibliography{./main.bib}

\end{document}